\documentstyle[12pt]{article}

\renewcommand{\theequation}{\thesection.\arabic{equation}}

\newlength{\extraspace}
\newlength{\extraspaces}
\newcounter{dummy}

\newcommand{\baa}{
\addtocounter{equation}{1}
\setcounter{dummy}{\value{equation}}
\setcounter{equation}{0}
\renewcommand{\theequation}{\thesection.\arabic{dummy}\alph{equation}}
\begin{eqnarray}
\addtolength{\abovedisplayskip}{\extraspaces}
\addtolength{\belowdisplayskip}{\extraspaces}
\addtolength{\abovedisplayshortskip}{\extraspace}
\addtolength{\belowdisplayshortskip}{\extraspace}}

\newcommand{\eaa}{
\end{eqnarray}
\setcounter{equation}{\value{dummy}}
\renewcommand{\theequation}{\thesection.\arabic{equation}}}

\newcommand{\be}{\begin{equation}
\addtolength{\abovedisplayskip}{\extraspaces}
\addtolength{\belowdisplayskip}{\extraspaces}
\addtolength{\abovedisplayshortskip}{\extraspace}
\addtolength{\belowdisplayshortskip}{\extraspace}}
\newcommand{\ee}{\end{equation}}

\newcommand{\ba}{\begin{eqnarray}
\addtolength{\abovedisplayskip}{\extraspaces}
\addtolength{\belowdisplayskip}{\extraspaces}
\addtolength{\abovedisplayshortskip}{\extraspace}
\addtolength{\belowdisplayshortskip}{\extraspace}}
\newcommand{\ea}{\end{eqnarray}}

\newcommand{\bd}{\begin{displaymath}
\addtolength{\abovedisplayskip}{\extraspaces}
\addtolength{\belowdisplayskip}{\extraspaces}
\addtolength{\abovedisplayshortskip}{\extraspace}
\addtolength{\belowdisplayshortskip}{\extraspace}}
\newcommand{\ed}{\end{displaymath}}

\newcommand{\ban}{\begin{eqnarray*}
\addtolength{\abovedisplayskip}{\extraspaces}
\addtolength{\belowdisplayskip}{\extraspaces}
\addtolength{\abovedisplayshortskip}{\extraspace}
\addtolength{\belowdisplayshortskip}{\extraspace}}
\newcommand{\ean}{\end{eqnarray*}}

\newcommand{\newsection}[1]{
\vspace{15mm}
\pagebreak[3]
\addtocounter{section}{1}
\setcounter{equation}{0}
\setcounter{subsection}{0}
\setcounter{footnote}{0}
\begin{center}
{\Large \thesection. #1}
\end{center}
\nopagebreak
\medskip
\nopagebreak}

\newcommand{\startappendix}{
\renewcommand{\thesection}{\Alph{section}}
\setcounter{section}{0}}

\newcommand{\newsubsection}[1]{
\vspace{1cm}
\pagebreak[3]

\addtocounter{subsection}{1}
\noindent{ \sc \thesubsection. #1}
\nopagebreak
\vspace{2mm}
\nopagebreak}

\newcommand{\nonu}{\nonumber \\[.5mm]}

\newcommand{\deel}[2]{{\textstyle{#1 \over #2}}}
\newcommand{\hf}{{\textstyle{1\over 2}}}

\newcommand{\ie}{{\it i.e.}}
\newcommand{\eg}{{\it e.g.\ }}
\newcommand{\re}{\mbox{I}\!\mbox{R}}
\newcommand{\lha}{\left[}
\newcommand{\rha}{\right]}

\def\inbar{\,\vrule height1.5ex width.4pt depth0pt}
\font\rms=cmr12 at 12pt
\def\ce{\relax\ifmmode\mathchoice
{\hbox{$\inbar\kern-.3em{\rm C}$}}
{\hbox{$\inbar\kern-.3em{\rm C}$}}
{\lower.9pt\hbox{\rms $\inbar\kern-.3em{\rm C}$}}
{\lower1.2pt\hbox{\rms $\inbar\kern-.3em{\rm C}$}}
\else{$\inbar\kern-.3em{\rm C}$}\fi}
\font\cmss=cmss12 \font\cmsss=cmss12 at 12pt
\def\ze{\relax\ifmmode\mathchoice
{\hbox{\cmss Z\kern-.4em Z}}{\hbox{\cmss Z\kern-.4em Z}}
{\lower.9pt\hbox{\cmsss Z\kern-.4em Z}}
{\lower1.2pt\hbox{\cmsss Z\kern-.4em Z}}\else{\cmss Z\kern-.4em Z}\fi}

\newcommand{\dif}{\partial}
\newcommand{\dbar}{\bar{\dif}}
\newcommand{\bnab}{\nabla_{\bar{z}}}
\newcommand{\hb}{\bar{h}}
\newcommand{\gb}{\bar{g}}
\newcommand{\bt}{\bar{T}}
\newcommand{\blam}{\bar{\Lambda}}
\newcommand{\beps}{\bar{\epsilon}}
\newcommand{\bz}{\bar{z}}

\newcommand{\abar}{A_{\bar{z}}}

\newcommand{\jbar}{J_{\bar{z}}}

\newcommand{\gam}{\Gamma}

\newcommand{\tr}{\mbox{Tr}}
\newcommand{\del}{\delta}
\newcommand{\pri}{\Pi_i}
\newcommand{\prk}{\Pi_k}
\newcommand{\prid}{\Pi_i^{\dagger}}
\newcommand{\prkd}{\Pi_k^{\dagger}}
\newcommand{\pry}{\Pi_y}

\newcommand{\vp}{\varphi}

\newcommand{\om}{\omega}
\newcommand{\bom}{\bar{\omega}}

\newcommand{\mub}{\bar{\mu}}

\newcommand{\bg}{{\bf g}}
\newcommand{\act}[1]{\deel{k}{2\pi}\int d^2z \, \tr ( {#1} ) }
\newcommand{\ac}[1]{\deel{1}{2\pi}\int d^2z \, \tr ( {#1} ) }
\newcommand{\actie}[1]{\deel{1}{2\pi}\int d^2z \, }

\newcommand{\vars}[2]{{\del {#1} \over \del {#2}}}

\newcommand{\mat}[9]{\left( \begin{array}{ccc}
				#1 & #2 & #3 \\
				#4 & #5 & #6 \\
				#7 & #8 & #9
                             \end{array} \right) }

\newcommand{\mats}[4]{\left( \begin{array}{cc}
				#1 & #2 \\
				#3 & #4
                             \end{array} \right) }

\newcommand{\cs}{Chern--Simons theory}

\newcommand{\maxdeg}{\mbox{\rm maxdeg}}
\newcommand{\diag}{\mbox{\rm diag}}
\newcommand{\mindeg}{\mbox{\rm mindeg}}

\newcommand{\pijl}{\stackrel{D'}{\longrightarrow}}

\newcommand{\np}[1]{Nucl. Phys. {\bf B#1}}
\newcommand{\cmp}[1]{Comm. Math. Phys. {\bf #1}}
\newcommand{\intmod}[1]{Int. Journal of Mod. Phys. {\bf A#1}}
\newcommand{\plb}[1]{Phys. Lett. {\bf B#1}}

\newcommand{\ad}[1]{\mbox{\rm ad}_{#1}}
\newcommand{\add}{\mbox{\rm ad}}
\newcommand{\aad}{\mbox{\rm Ad}}
\newcommand{\im}{\mbox{\rm im}}
\newcommand{\bn}{{\bf 0}}
\newcommand{\eps}{\epsilon}
\newcommand{\End}{\mbox{\rm End}}

\begin{document}
\addtolength{\baselineskip}{.7mm}

\thispagestyle{empty}
\begin{flushright}
{\sc THU}-92/14\\
6/92
\end{flushright}
\vspace{1.5cm}
\setcounter{footnote}{2}
\begin{center}
{\LARGE\sc{Covariant $W$ Gravity \& its Moduli Space}}\\[4mm]
{\LARGE\sc{from Gauge Theory}}\\[2cm]

\sc{Jan de Boer\footnote{e-mail: deboer@ruunts.fys.ruu.nl}
and Jacob Goeree\footnote{e-mail: goeree@ruunts.fys.ruu.nl}}
\\[8mm]
{\it Institute for Theoretical Physics\\[2mm]
University of Utrecht\\[2mm]
Princetonplein 5\\[2mm]
P.O. Box 80.006\\[2mm]
3508 TA Utrecht}\\[1.5cm]

{\sc Abstract}\\[1cm]
\end{center}

\noindent
In this paper we study arbitrary $W$ algebras related to
embeddings of $sl_2$ in a Lie algebra ${\bf g}$. We will
give a simple general formula for all $W$ transformations,
which will enable us to construct the covariant action for
general $W$ gravity. It turns out that this covariant action
is nothing but a Fourier transform of the WZW action. The
same general formula provides a `geometrical' interpretation
of $W$ transformations: they are just a homotopy contraction
of ordinary gauge transformations. This is used to
argue that the moduli space relevant to $W$ gravity is part
of the moduli space of $G$-bundles over a Riemann surface.
\vfill

\newpage

\newsection{Introduction}

One of the most dramatic discoveries concerning quantum gravity in
two dimensions has been Polyakov's observation that 2D
gravity in the light-cone (or chiral) gauge has a $Sl(2,\re)$ gauge
symmetry \cite{pol2}. The reason for the occurrence of this unexpected
symmetry has become more clear from Polyakov's
`soldering procedure' \cite{pol3}, which shows how two dimensional
diffeomorphisms can be obtained from $Sl(2,\re)$ gauge transformations.
This procedure strongly hints at a connection between the gravitational
action and the WZW action for a $Sl(2,\re)$ gauge field, something
which was indeed established by \cite{pol3,beroog}, who showed
that the gravitational action is related to the WZW action of a
{\em constrained}\, $Sl(2,\re)$ gauge field. The consequences of these
results are quite far-reaching: $i)$ they relate the correlators in
chiral gravity to those in $Sl(2,\re)$ gauge theory, $ii)$ they pave the
way to study renormalization effects in 2D quantum gravity \cite{kpz},
and $iii)$ replacing $Sl(2,\re)$ by some other
non-compact real Lie group they open the possibility to
study higher spin generalizations of ordinary
2D gravity, \ie\ $W$ gravity theories.

In this paper we will concern ourselves with the last point. We will
show that many of the results mentioned above for the case of ordinary
gravity can be extended to the case of $W$ gravity. The $W$ algebras
we will consider are quite general: they are related to $sl_2$
embeddings in a Lie algebra $\bg$ \cite{adam}, which among others contain
the standard $W_N$ algebras \cite{sammy}.
For these generalized $W$ algebras we
will construct the chiral and covariant action, and show that {\em both}\,
can be understood in terms of a WZW model based on the Lie group
${\bf G}$ corresponding to $\bg$. For the chiral action this implies that
correlation functions in chiral $W$ gravity can be computed from
the operator product expansion of the Kac--Moody
currents of the WZW theory based on ${\bf G}$.

More importantly, having a clear understanding of
the {\em covariant}\, action for $W$ gravity gives us information
about the degrees of freedom of $W$ gravity.
To appreciate this point, consider ordinary
gravity. The metric in ordinary gravity has three degrees of
freedom, the Liouville field $\phi$ and a complex Beltrami differential
$\mu$. These parametrize a general metric via $ds^2=e^{-2\phi}|dz+
\mu d\bar{z}|^2$. Now imagine that we start working in the chiral gauge
$\mub=\phi=0$ and construct the chiral gravitational action $\gam[\mu]$,
and similar, working in the gauge $\mu=\phi=0$, its partner of opposite
chirality $\gam[\mub]$. Suppose we want to couple
these chiral actions in such a way that the resulting
theory is invariant under diffeomorphisms.
For ordinary gravity we know how to do
this: we have to introduce the Liouville field $\phi$, and build
out of this $\phi$ and $\mu,\mub$ a metric as indicated above.
Given this metric we construct an action using only
covariant quantities, which is such that it reduces to the
chiral action $\gam[\mu]$ or $\gam[\mub]$
if one takes the corresponding gauge.
Since only covariant quantities are used,
this action is then diffeomorphism invariant
by construction. For $W$ gravity however, this procedure is not
as clear cut as this. We can construct $\gam[\mu_i]$,
the $W$ analogue of the chiral action $\gam[\mu]$, but
it is not at all clear which fields we should introduce in
order to construct out of these and the $\mu_i,\mub_i$ the
$W$ analogue of the metric, and possible higher spin fields,
simply since we have no idea what these $W$ analogues are.
But if we somehow can circumvent this problem of not knowing what
the $W$ analogue of the metric is, and still construct a covariant
action, the extra fields one has to introduce for $W$ gravity should
arise automatically. In \cite{kj2}, such extra fields were
found necessary to construct an action of matter covariantly
coupled to the $W_3$ algebra.

This is in fact what we shall do in this paper. We will show that
the covariant action for $W$ gravity can be obtained by Fourier
transformation of the WZW action based on ${\bf G}$. This Fourier
transformation will turn out to be easy and exact to do, since the
integrals involved are simple Gaussian integrals. The resulting
action is closely related to the action of a gauged WZW theory
\cite{gwzw}, albeit that the subgroup one usually gauges is
replaced by the $W$ algebra.  This procedure
will make it clear that the extra fields we have to introduce
for $W$ gravity can be labeled by one extra group variable $G\in {\bf
G}$. Specializing to a particular $W$ algebra one can subsequently show
that some of the degrees of freedom labeled by $G$ are non-propagating
and can therefore be integrated out \cite{jj2}. For example for ordinary
gravity it will turn out that only the Cartan subgroup of
${\bf G}=Sl(2,\re)$ labels a true degree of freedom,
which is of course nothing
but the Liouville field $\phi$. For other $W$ algebras one is
generically left with the Cartan subgroup of ${\bf G}$ plus more.
Working out the action for these $W$ algebras one finds that the
Cartan subgroup of ${\bf G}$ gives rise to a Toda system coupled
in some way to the parameters $\mu_i,\mub_i$ of the $W$ algebra,
something which might have been expected from the intimate
relation between Toda theories and $W$ algebras \cite{toda}.
But since the degrees of freedom for generic $W$ algebras
are not labeled by the Cartan subgroup only, this Toda system
will in general be supplemented by some extra piece.
This structure of the covariant action was previously found in
\cite{jj2}, where we studied $W$ algebras using \cs. The advantage
of our present method is that: $i)$ it gives better insight in what
the degrees of freedom for $W$ gravity are, $ii)$ it is also
applicable to non-standard $W$ algebras, $iii)$ it gives a
relation between the partition function for $W$ gravity and the
partition function for WZW theory, and $iv)$ the constructed
covariant action is manifestly invariant under left
and right $W$ transformations throughout the construction.

Another subject we will concern ourselves with in this paper
regards the `moduli space for $W$ gravity'.
Recall that a nontrivial ingredient of string theory in
general and of quantum gravity in particular is that the space of
metrics modulo Weyl transformations and diffeomorphisms is not
trivial, but constitute the well known moduli space of Riemann
surfaces. For ordinary gravity this space is well understood, but
what the appropriate generalization of this space to $W$ gravity is,
is still unclear. Topological field theory and the
matrix model approach to 2D gravity seem to
suggest that the moduli space for $W$ gravity is somehow related to
the moduli space of flat $Sl(N,\re)$ bundles \cite{li1}.
In this paper we will show that with an appropriate definition of
$W$ moduli space (or more precisely $W$ Teichm\"uller space),
which essentially defines it to be the space of $W$
fields modulo $W$ transformations, this space can be
computed for arbitrary $W$ gravity. It turns out that for the
standard $W_N$ algebras, this space is a component of the moduli
space of flat $Sl(N,\re)$ bundles (as conjectured by E. Witten in
\cite{li2}), and that for the other $W$ algebras it is a space
whose interpretation remains unclear.

The computation involves several steps that are of independent
interest. First of all, a formulation of $W$ algebras on arbitrary
Riemann surfaces is given, thereby generalizing \cite{ward}.
This is necessary, because the existence of moduli is intimately
related to the fact that the Riemann surface has nontrivial
topology. This formulation involves certain nontrivial $Sl(N,\ce)$
bundles over the Riemann surface, forcing us to extend the
definition of the WZW action to the case of nontrivial bundles.
This will enable us to write down chiral and covariant actions on
arbitrary genus Riemann surfaces. Next, we show how one can obtain
$W$ transformations from ordinary gauge transformations by a
mathematical procedure called homotopy contraction, providing an
alternative geometrical picture of $W$ transformations. To complete
the proof of the relation between $W$ moduli space and flat
$Sl(N,\re)$ bundles, we show that the $W$ moduli space is in a {\it
natural}\, way a subspace of the so-called moduli space of Higgs
bundles \cite{hitchin,simpson}. This generalizes the relation
between metrics on a Riemann surface and Higgs bundles previously
discussed by Hitchin.

This paper is organized as follows: in section 2 we show how the
covariant action for ordinary gravity can be obtained from
$Sl(2,\re)$ WZW theory. In section 3 we review the construction of
general $W$ algebras associated to $sl_2$ embeddings in certain
non-compact Lie algebras $\bg$, and present a general formula for
$W$ transformations using an operator $L$ that depends only on the
$sl_2$ embedding. This general formulation is used in section 4 to
construct the chiral and covariant actions for general $W$ gravity.
Finally, in section 5 we discuss the generalization of all this to
higher genus Riemann surfaces and compute the moduli space for $W$
gravity.

In this paper we will only consider classical $W$ algebras, by
which we mean Poisson algebras that describe the large $c$
behavior of quantum $W$ algebras, \ie\ of all numerical
coefficients of the quantum $W$ algebras we only keep the highest
power of $1/c$. In the literature, one sometimes finds different
definitions of a classical $W$ algebra, that are further reduced
versions of the $W$ algebras we consider. All our computations will
also be valid only to lowest order in $1/c$.

\newsection{Induced Gauge Theory and Gravity in Two Dimensions}

Consider some action $S(\mbox{matter},g_{ab})$, describing a
set of matter fields coupled to gravity, which is invariant
under diffeomorphisms and Weyl transformations.
Integrating out the matter from such a theory one
obtains a gravitational induced action $\Gamma[g_{ab}]$.
If the theory has no anomalies, $\Gamma[g_{ab}]$
reduces to an action on moduli space, since on the
classical level the number of degrees of freedom of the metric
equals the number of invariances. However, as is well known,
the procedure of integrating out the matter cannot be
done in both a Weyl and diffeomorphism invariant way,
leading to a non-trivial $g$ dependence of $\Gamma[g_{ab}]$.
Using a diffeomorphism invariant regulator to handle
the matter integration one obtains the following
expression for the induced gravitational action:
\be \label{polyak}
\Gamma[g_{ab}]=\frac{c_m}{96\pi} \int R \, \frac{1}{\Box} R,
\ee
a result first obtained by Polyakov \cite{pol1}. Note
that $c_m$, the central charge of the matter system, is the
only remnant of the matter system we used to define the
induced action.

In this paper we will concern ourselves with the generalization
of the above result for ordinary gravity
to the case of $W$ gravity. At first sight this seems rather
difficult, since up till now it is by no means clear what
the $W$ generalizations of the metric, diffeomorphisms, etc. are.
We will circumvent these difficulties by first relating
(\ref{polyak}) to the action of a $Sl(2,\re)$ WZW model.
Then the generalization to $W$ gravity will consist of
replacing $Sl(2,\re)$ by some other non-compact real Lie
group.

The fact that $\Gamma[g_{ab}]$ is somehow related to
$Sl(2,\re)$ WZW theory, might have been expected from the
works of Polyakov \cite{pol2} and Bershadsky and Ooguri
\cite{beroog}, who showed that a similar relation exists for
induced gravity in the chiral gauge\footnote{By this we mean
that a general metric, which can always be written as
$ds^2=e^{-2\vp}|dz+\mu d\bar{z}|^2$, is reduced to
$ds^2=dzd\bar{z}+\mu d\bar{z}d\bar{z}$, so we impose the gauge
$\vp=\mub=0$.}. Indeed it was Polyakov who first discovered that
two dimensional gravity in the chiral gauge can be described
in terms of a $Sl(2,\re)$ current algebra \cite{pol2}.
The reason for the appearance of this $Sl(2,\re)$ current
algebra was further elucidated in \cite{beroog},
where it was shown that the partition function for gravity in the
chiral gauge can be written in terms of a constrained
WZW model:
\be \label{partfun}
\int D\mu \exp(-\gam[\mu])=\int \frac{Dg}{\mbox{(gauge vol)}}
\delta(J_z^+-1)\exp(-kS^-_{wzw}(g)),
\ee
where $J_z=g^{-1}\dif g$, $k$ is related to the matter central charge
$c_m$ via $c_m \sim -6k$, and we have to divide by the volume
of the Borel subgroup $B^-=\exp(\epsilon T^-)$ under which the
constrained WZW model is invariant. This result can be straightforwardly
generalized to the case of {\em chiral}\, $W_N$ gravity (at least
in the large $k$ limit) by considering a constrained $Sl(N,\re)$
model instead of the $Sl(2,\re)$ model.

We will take a different approach to the subject here, and
study the {\em covariant}\, induced gravity theory.
We will argue that also the partition function for
this theory can be understood in terms of
a $Sl(2,\re)$ WZW model. The correspondence we will
arrive at reads:
\be \label{corr2}
\int \frac{Dg_{ab}}{\mbox{(gauge vol)}} \exp(-\Gamma[g_{ab}]) =
\int \frac{DgDGD\gb}{\mbox{(gauge vol)}}
\delta(J_z^+-1) \delta(\bar{J}^-_{\bar{z}}-1)
\exp (-kS^{-}_{wzw}(gG\gb^{-1})),
\ee
where on the l.h.s. we have to divide by the volume of the
diffeomorphism group and on the r.h.s. we have to divide by the
symmetry group of the left, right constrained WZW model.
Also this formula can be easily generalized to the
case of general $W$ gravity by replacing $Sl(2,\re)$ by some other
non-compact real Lie group, something which will be done in
section~4.

Before we come to the above result (\ref{corr2}),
we will first explain how chiral induced gravity is related
to chiral induced $Sl(2,\re)$ gauge theory. We start by
reviewing some generalities of chiral induced gauge theories.
Next we show how the defining relation for $\Gamma[\mu]$
can be obtained from $Sl(2,\re)$ gauge theory,
which makes it possible
to obtain an expression for $\Gamma[\mu]$. Then, to obtain the
covariant action for induced gravity, we couple the left and
right sector, \ie\ $\Gamma[\mu]$ and $\Gamma[\mub]$,
in such a way that the resulting theory is invariant under
diffeomorphisms, leading to (\ref{corr2}).

\newsubsection{Chiral Induced Gauge Theories}

Given some action $S(\mbox{matter},\abar)$, describing a
chiral gauge field $\abar$ coupled to some matter system,
which is invariant under gauge transformations,
we define the induced action for chiral gauge theory as:
\ba \label{indact}
\exp(-\gam[\abar]) &=& \int D(\mbox{matter})
\exp(-S(\mbox{matter},\abar)) \nonu
&=& \left< \exp-\ac{\abar J_z} \right>,
\ea
where $J_z^a$ are the Noether currents corresponding to the gauge
symmetry. The operator product expansion
\be \label{ope1}
J_z^a(z)J_z^b(w) \sim \frac{f^{ab}_c}{z-w}J_z^c(w)
+\frac{k}{(z-w)^2}\eta^{ab},
\ee
where $k$ is the level of the matter current algebra, implies
the following differential equation for the induced action:
\be \label{difeq}
\left( \eta^{ab}\dbar+f^{ab}_c\abar^c(z,\bz)\right)
\vars{\gam[\abar]}{\abar^b(z,\bz)}=\frac{k}{2\pi}\dif
\abar^a(z,\bz).
\ee
The solution to this equation is well known \cite{beroog}:
it is given by $\gam[\abar]=kS^{+}_{wzw}(g)$, where $\abar$ is
related to $g$ via $\abar=g^{-1}\dbar g$\footnote{Such a
parametrization for $\abar$ is of course only generic if we are
working on the plane, as we do here. The generalization of
this to higher Riemann surfaces will be dealt with
in section~5.}. So starting with any
matter system coupled to a chiral gauge field $\abar$, the
induced gauge theory action will always be given by
$kS^{+}_{wzw}(g)$, with $k$ the level of the matter current
algebra. Note that if one defines
\be \label{indcur}
J_z^{ind}=\frac{2\pi}{k}\vars{\gam[\abar]}{\abar},
\ee
(\ref{difeq}) states that the pair $\{ J_z^{ind},\abar \}$
has vanishing curvature. In this way the WZW functional
can be viewed upon as the solution of the zero-curvature condition.

In the definition of $\gam[\abar]$ the gauge field $\abar$
is treated as a classical background field.
To `quantize' the gauge field we consider the generating
functional for $\abar$ correlators:
\be \label{genfun}
\exp(-\gam[U_z])=\int D\abar \exp\left(-\gam[\abar]+
\act{\abar U_z}\right).
\ee
$\gam[U_z]$, which can be seen as the Fourier transform
of $\gam[\abar]$, can be computed explicitly if one uses
the Polyakov--Wiegman identity for the WZW action (see the Appendix):
\be \label{PW}
S^{+}_{wzw}(gh)=S^{+}_{wzw}(g)+S^{+}_{wzw}(h)+
\deel{1}{2\pi}\int d^2z\,\tr(g^{-1}\dbar g \dif h h^{-1}).
\ee
For this, parametrize $\abar=g^{-1}\dbar g$ and $U_z=h^{-1}\dif
h$, and replace the measure $D\abar$ in (\ref{genfun}) by the
Haar measure $Dg$\footnote{Here we neglect the Jacobian
$| \frac{D\abar}{Dg}|$
arising from the variable transformation, which can be
computed exactly as $\mbox{det}(\dbar+\abar)=\exp(2c_VS^{+}_{wzw}(g))$,
where $c_V$ is the dual Coxeter number.
This contribution can be ignored since we take the large $k$
limit.}. One has:
\be \label{genfun2}
\exp(-\gam[U_z])=\exp(kS^{-}_{wzw}(h)) \int Dg
\exp(-kS^{+}_{wzw}(gh^{-1})).
\ee
Since the Haar measure is invariant under the action of the
gauge group, the result of the $g$ integration is independent of
$h$, and we obtain $\gam[U_z]=-kS^{-}_{wzw}(h)$. This result could
also have been derived as follows: since in the large $k$
limit the semi-classical approximation to the path integral in
(\ref{genfun}) is exact, we solve for the stationary points of
$\gam[\abar]-\act{\abar U_z}$. Using the results of Appendix~A
one easily verifies that these are given
by $g=h$. Plugging this back into (\ref{genfun}) results in
$\gam[U_z]=-kS^{-}_{wzw}(h)$.

In the next subsection we will consider the analogous
case of induced chiral gravity.
The operator product expansion of the matter stress energy tensor
$T_{mat}$ gives a differential equation for the
induced action $\exp(-\gam[\mu])=\left< \exp(-\deel{1}{2\pi}\int \mu
T_{mat})\right>$ quite similar to (\ref{difeq}).
We will show that this differential equation
can be obtained from the zero-curvature condition for a
$Sl(2,\re)$ gauge field if we constrain the gauge field
to a particular form. (The origin of these constraints can be
traced back to the work of Drinfel'd and Sokolov \cite{ds},
who showed how such a constrained gauge field in a natural
way (via Hamiltonian reduction) gives rise to the second Gelfand-Dickii
bracket \cite{gelfdick}, which
exactly reproduces the classical form of $W$ algebras.)
Given this relation between the defining differential equation for
$\gam[\mu]$ and the zero-curvature condition for a constrained gauge
field, it will be straightforward
to write down an expression for $\gam[\mu]$. Before we come to
this, we will spend a few words on how two dimensional diffeomorphisms
follow from $Sl(2,\re)$ gauge transformations if one considers
constrained gauge fields, a construction due to Polyakov \cite{pol3}.

\newsubsection{Diffeomorphisms from Gauge Transformations}

Consider a constrained $Sl(2,\re)$ gauge potential $A_z$ of the form
\be \label{restr}
A_z=\Lambda+W=\mats{0}{1}{T}{0},
\ee
where $\Lambda$ is the constant matrix $\mats{0}{1}{0}{0}$,
and $W$ is the matrix containing $T$, the only dynamical component
of $A_z$. We will focus on gauge transformations that leave
the form of $A_z$ invariant. As we will see in a moment, finding
such gauge transformations is equivalent to finding a current
$\jbar$, such that the pair $\{ A_z, \jbar \}$ has
vanishing field strength.
One can easily solve for two of the three components of $\jbar$
from the zero-curvature equation
\be \label{zerocurv}
F(A_z,\jbar)=\dif \jbar - \dbar A_z +\lha A_z,\jbar \rha =0.
\ee
The result is that $\jbar$ should be of the form
\be \jbar=\mats{\hf \dif \mu}{\mu}{\mu T - \hf \dif^2 \mu}{-\hf
\dif \mu}. \label{defabar}
\ee
Substituting this in (\ref{zerocurv}) leaves us with one
equation, which is precisely the chiral Virasoro Ward-identity
\be \label{Virasoro}
\left(\dbar-\mu\dif-2(\dif\mu)\right)T + \hf\dif^3\mu=0.
\ee
If we replace in this equation $\mu$ by $\epsilon$ and $\dbar T$
by $\delta_{\epsilon}T$, we find precisely the transformation
rule for an energy-momentum tensor with $c=6$ under a general
co-ordinate transformation. Going back to (\ref{zerocurv}), we
see that
\be
\delta_{\epsilon} A_z = D_{A_z} (\jbar(\epsilon)),
\ee
where $D_{A_z}=\dif + [A_z, .\, ]$ is the covariant derivative,
and $\jbar(\epsilon)$ is (\ref{defabar}) with $\mu$ replaced
by $\epsilon$.
Note that $\jbar(\epsilon)$ is indeed the
unique gauge transformation that leaves the form of $A_z$
invariant. This shows that co-ordinate transformations take
the form of field dependent gauge transformations. This
generalizes to arbitrary $W$ algebras as will be discussed
in section 3.

\newsubsection{Chiral Gravity from $Sl(2,\re)$ Gauge Theory}

As emphasized in \cite{pol3} the above geometrical observation
hints at a connection between the gravitational action and
the action for induced $Sl(2,\re)$ gauge theory. Consider the
induced action $\exp(-\gam[\mu])=\left< \exp(-\deel{1}{2\pi}
\int \mu T_{mat}) \right>$ for chiral gravity. Since $\gam[\mu]$
is the generating functional for the correlation functions of the
matter stress energy tensor $T_{mat}$, the operator product expansion
\be \label{Tope}
T_{mat}(z)T_{mat}(w) \sim\frac{c_m/2}{(z-w)^4}+
\frac{2T_{mat}(w)}{(z-w)^2}+ \frac{\dif T_{mat}(w)}{z-w},
\ee
implies the following differential equation for $\gam[\mu]$:
\be \label{difeq2}
\left( \dbar -\mu \dif -2(\dif \mu) \right)\vars{\gam[\mu]}{\mu}=
\frac{c_m}{24\pi}\dif^3 \mu.
\ee
Defining
\be \label{deftind}
T_{ind}=\frac{2\pi}{k}\vars{\gam[\mu]}{\mu},
\ee
(where we made the identification $c_m\sim -6k$)
we see that this defining relation for $\gam[\mu]$ is precisely
the same as the chiral Virasoro Ward identity (\ref{Virasoro}),
obtained in the previous subsection by considering
the zero-curvature condition for a constrained
$Sl(2,\re)$ gauge field.

As explained in section~2.1, the solution to this zero-curvature
condition is given by the WZW functional.
This implies that $\gam[T]$, which is the Fourier transform of
$\gam[\mu]$, is simply given by $\gam[T]=-kS^{-}_{wzw}(g)$,
where $g$ is such that $g^{-1}\dif g=\Lambda + W$ (see (\ref{restr})).
In turn, $\gam[\mu]$ can then be determined by Fourier
transformation:
\ba \label{ft}
\exp(-\gam[\mu])&\!\!=\!\!&\int DT \exp\left( -\gam[T]-\deel{k}{2\pi}\int
d^2z\,\mu T\right) \\
&\!\!=\!\!&\int DTDh\exp\left(-kS^{+}_{wzw}(h)+\deel{k}{2\pi}
\int d^2z\, \tr(g^{-1}\dif g h^{-1}\dbar h)-
\deel{k}{2\pi}\int d^2z\,\mu T\right) \nonu
&\!\!=\!\!&\int Dh\, \delta(\jbar^{+}-\mu)\exp\left(
-kS^{+}_{wzw}(h)+\act{\Lambda h^{-1}\dbar h}\right), \nonumber
\ea
where $\jbar=h^{-1}\dbar h$, and we used the Polyakov--Wiegman
identity for the WZW action to go from the first to the second line.
The `extra' term in the exponent, $\tr(\Lambda h^{-1}\dbar h)$,
was previously found in \cite{pol3} and in a different form in
\cite{kj1}.
Before we come to our final, more manageable result for
$\gam[\mu]$ note that the argument of the exponent in the last
line can be written as $-kS^{+}_{wzw}(hf)$, with
$f=\exp(-z\Lambda)$. Using the invariance of the Haar measure
we thus find
\be \label{expl}
\exp(-\gam[\mu])=\int
Dh\,\delta(\mu-(\jbar^{+}+2z\jbar^{0}-z^2\jbar^{-}))
\exp(-kS^{+}_{wzw}(h)),
\ee
which explains the occurrence of the $Sl(2,\re)$ current algebra
in chiral induced gravity.

In order to obtain a more tangible expression for $\gam[\mu]$
we take the large $k$ limit of (\ref{ft}). In this limit
$\gam[\mu]$ is given by the semi-classical approximation to
(\ref{ft}). As explained in Appendix~A, the
equation of motion for $h$ has as its solution that
$h^{-1}\dbar h$ should be of the form (\ref{defabar}).
So $\gam[\mu]=kS^{+}_{wzw}(h)-\act{\Lambda h^{-1}\dbar h}$,
with $\jbar=h^{-1}\dbar h$ of the form (\ref{defabar}).
In (\ref{defabar}) $T$ is related
to $\mu$ via the chiral Virasoro Ward identity (\ref{Virasoro}).
This identity can in principle be used to determine $T$ as a
{\em non-local}\, function of $\mu$. In the next subsection we
will see how a {\em local}\, expression for $T$
can be obtained when one takes a convenient
parametrization for $\mu$ \cite{pol2}.

\newsubsection{Local Expressions for the Induced Actions
$\gam[T]$ and $\gam[\mu]$}

We saw that the induced action $\gam[\mu]$ can be written
in terms of a WZW model, in which the group element $h$ is constrained
such that $\jbar=h^{-1}\dbar h$ is of the particular form
(\ref{defabar}). To find a
more explicit form for the induced action we now want to solve the
constraints imposed on $h$. For this we take the following
Gauss decomposition for $h$:
\be \label{gauss1}
h=\mats{1}{F}{0}{1}\mats{e^{\psi}}{0}{0}{e^{-\psi}}
\mats{1}{0}{V}{1}.
\ee
Demanding $\jbar$ to be of the form (\ref{defabar}) gives the following
identities for $F,V,\psi$: $\psi=\hf \log \dif F$ and $V=
-\dif \psi$, and $F$ the only independent variable is related
to $\mu$ through $\mu=\dbar F /\dif F$. From (\ref{defabar}) one
obtains the following {\em local}\, expression for $T$:
\be \label{Sder}
T = \hf \{ F,z \} = \frac{1}{2} \left( \frac{\dif^3 F}{\dif F}-
\frac{3}{2} \left( \frac{\dif^2 F}{\dif F}\right)^2\right),
\ee
\ie\ the Schwarzian derivative of $F$.
One easily verifies that this $\mu$ and $T$ indeed solve the
chiral Virasoro Ward identity (\ref{Virasoro}).
Inserting this solution for $h$ back into the result
$\gam[\mu]=kS^{+}_{wzw}(h)-\act{\Lambda h^{-1}\dbar h}$
we found previously, gives \cite{pol2}
\be \label{localex}
\gam[\mu(F)]=\deel{k}{8\pi}\int d^2z\,\frac{\dbar F}{(\dif F)^3}
\left( \dif^3 F\dif F-(\dif^2 F)^2 \right).
\ee

In a similar way we can find a local expression for
$\gam[T]=-kS^{-}_{wzw}(g)$, with $g^{-1}\dif g$ restricted to be
of the form (\ref{restr}). Taking the above Gauss decomposition
for $g$ we find that the constraints on $g$ are also solved by
$\psi=\hf \log \dif F$, $V=-\dif \psi$. Again $T$ is the Schwarzian
derivative of $F$, and $\gam[T]$ becomes
\be \label{localex2}
\gam[T(F)]=-\deel{k}{8\pi}\int d^2z\,\frac{\dbar F}{(\dif F)^3}
\left( \dif^3 F \dif F -2 (\dif^2 F)^2\right).
\ee
Note that $\gam[T(F)]=-\gam[\mu(F^{-1})]$, where $F^{-1}$
is such that $F(F^{-1}(z,\bz),\bz)=z$. So in this picture Fourier
transformation just amounts to taking the inverse function.


\newsubsection{Covariant Gravity from $Sl(2,\re)$}

Previously we demonstrated how diffeomorphisms arise from $Sl(2,\re)$
gauge transformations if the gauge field is restricted to a particular
form. This fact enabled us to compute the action for induced chiral
gravity
\be \label{res1}
\gam[\mu] = kS^{+}_{wzw}(h)-\act{\Lambda h^{-1}\dbar h},
\ee
where $\jbar=h^{-1}\dbar h$ is restricted as in (\ref{defabar}).
As explained diffeomorphisms are represented by gauge transformations:
$\delta_{\epsilon}h=hX(\epsilon)$, where $X(\epsilon)$
is (\ref{defabar}) with $\mu$
replaced by $\epsilon$. Note that this transformation rule for $h$
reproduces the standard transformation rule for the
Beltrami differential $\mu$: $\delta_{\epsilon}\mu=
\dbar \epsilon+\epsilon\dif \mu-\mu \dif \epsilon$.
Given the behavior of the WZW functional under gauge transformations,
one easily verifies the following `anomalous' transformation
rule of the induced action: $\delta_{\epsilon}\gam[\mu]=
-\deel{k}{4\pi}\int d^2z\,\epsilon\dif^3 \mu$.

Instead of gauging the right handed sector which
we did up till now, we could as well gauge the left handed sector. Of
course, all the results derived above for the former gauge have their
analogs for this new gauge. The induced action $\gam[\mub]$ is given
by
\be \label{dualac}
\gam[\mub]=kS^{-}_{wzw}(\hb)-\act{\blam \hb^{-1}\dif \hb},
\ee
where $\hb$ is such that $\hb^{-1}\dif \hb$ is given by the transpose
of (\ref{defabar}) with $\mu,T$ replaced by $\mub,\bt$, and $\blam$
is the transpose of $\Lambda$.
Similarly, diffeomorphisms are represented by $\delta_{\beps}\hb=\hb
\bar{X}(\beps)$, where $\bar{X}(\beps)$ is the transpose of $X(\eps)$
with $\eps$ replaced by $\beps$. Under diffeomorphisms
$\gam[\mub]$ transforms as:
$\delta_{\beps}\gam[\mub]=-\deel{k}{4\pi}\int d^2z\,\beps \dbar^3\mub$.

The main goal of this section is to construct out of the chiral
actions $\gam[\mu]$ and $\gam[\mub]$ a covariant action
$S_{cov}(\mu,\mub,\Phi)$ which is invariant under
diffeomorphisms. Here $\Phi$ denotes some set of additional
fields which may be needed to make the action invariant.
Stated differently, we are looking for a `local counterterm' $\Delta
\gam[\mu,\mub,\Phi]$ whose anomalous behavior under
diffeomorphisms exactly cancels that of the chiral actions. Then the
covariant action will be given by
$S_{cov}=\Delta \gam[\mu,\mub,\Phi]+\gam[\mu]+\gam[\mub]$.
An interesting consequence of this decomposition of $S_{cov}$ into
a left and right chiral action and a mixed term $\Delta \gam$ is
that if we define the wave-function $\Psi[\mu]=\exp(-\gam[\mu])$,
the partition function for induced gravity can be
written as the norm squared of $\Psi[\mu]$:
\be \label{ip}
{\cal Z}_{grav}\equiv \int \frac{Dg_{ab}}{\mbox{(gauge vol)}}
\exp(-\gam[g_{ab}])=\left< \Psi,\Psi \right>,
\ee
where (gauge vol) denotes the volume of the diffeomorphism group,
and the inner-product is defined as:
\be \label{defip}
\left< \Psi_1,\Psi_2 \right> \equiv \int
\frac{D\mu D\mub D\Phi}{\mbox{(gauge vol)}}
\,\exp \Bigl(-\Delta \gam[\mu,\mub,\Phi]\Bigr)
\,\Psi_1[\mub]\,\Psi_2[\mu].
\ee
{}From standard Fourier theory we then know that the partition function
can also be written as the norm squared of the Fourier transformed
wave-function $\Upsilon[T]=\exp(-\gam[T])$, for which the inner-product
is given by:
\be \label{defip2}
\left< \Upsilon_1,\Upsilon_2 \right> \equiv \int
\frac{DT D\bt D\Phi}{\mbox{(gauge vol)}}
\,\exp \left(-\Delta \gam[T,\bt,\Phi]\right)
\,\Upsilon_1[\bt]\,\Upsilon_2[T],
\ee
where $\Delta \gam[T,\bt,\Phi]$ is the Fourier transformed
of $\Delta \gam[\mu,\mub,\Phi]$ with respect to $\mu,\mub$.

This second formulation is in fact more suitable,
because, as we will see shortly, the
covariant action $S_{cov}(T,\bt,\Phi)=\Delta \gam[T,\bt,\Phi]
+\gam[T]+\gam[\bt]$, is easier to derive.
Recall that $\gam[T]$ is simply given by a restricted WZW functional
and that diffeomorphisms are represented by the gauge transformations
$\delta_{\eps}g=gX(\eps)$.
Similarly, $\gam[\bt]=-kS^{+}(\gb)$,
with $\gb^{-1}\dbar \gb=\blam+\bar{W}$
(the transpose of (\ref{restr}) with $T$ replaced by
$\bt$), and diffeomorphisms are represented as $\delta_{\beps}\gb=
\gb\bar{X}(\beps)$. From all this it is obvious that if we introduce
as our extra field $\Phi$ one more group variable $G$, which under
diffeomorphisms transforms as:
\be \label{transforma}
\delta_{\eps,\beps}G=-X(\eps)G+G\bar{X}(\beps),
\ee
the covariant action we are looking for is
given by\footnote{Note
that $S_{cov}$ cannot just be any function of $gG\gb^{-1}$ since for
the chiral gauges $G={\bf 1},\bar{T}=0$ or $G={\bf 1},T=0$ the
covariant action should reduce to the corresponding chiral action.}
$S_{cov}(T,\bt,G)=-kS^{-}_{wzw}(gG\gb^{-1})$.

Using the Polyakov--Wiegman identity for the WZW functional,
one now easily
computes $\Delta \gam[T,\bt,G]$ to be:
\ba \label{delgamt}
\Delta \gam[T,\bt,G] &=&
\act{(\blam+\bar{W})G^{-1}\dif G}
-\act{(\Lambda+W)\dbar G G^{-1}} \nonu
& &+\act{(\Lambda+W)G(\blam+\bar{W})G^{-1}}-kS^{-}_{wzw}(G).
\ea
$\Delta \gam[\mu,\mub,G]$ can now be obtained from this
by Fourier transformation. In the chiral case this was a
non-trivial, although not impossible procedure, but here
things are much more simple since the integral over
$T,\bt$ is a Gaussian integral, and can thus easily
be carried out.
Indeed, if we parametrize $G$ by the Gauss decomposition:
\be \label{gauss2}
G=\mats{1}{0}{\om}{1}\mats{e^{\phi}}{0}{0}{e^{-\phi}}
\mats{1}{-\bom}{0}{1},
\ee
the saddle-point equations for $T,\bar{T}$ become
\ba \label{eqntt}
T &=& \om^2-\dif \om +\mub e^{-2\phi}, \nonu
\bar{T} &=& \bom^2-\dbar \bom +\mu e^{-2\phi}.
\ea
Substituting these expressions back leads to the following
result for $\Delta \gam[\mu,\mub,G]$ \cite{herman,jj2}
\ba \label{cosl2}
\Delta \gam[\mu,\mub,G]= \deel{k}{2\pi}\int d^2z\, \Bigl[
\dif \phi \dbar \phi + \omega(2\dbar \phi +\dif \mu)
+\bar{\omega} ( 2\dif \phi + \dbar \bar{\mu}) \nonu
\,\,\,\,+\mu \omega^2 + \bar{\mu}\bar{\omega}^2 +
2\omega\bar{\omega}-(1-\mu\bar{\mu})e^{-2\phi} \Bigr].
\ea
{}From (\ref{cosl2}) we recognize that $\om,\bom$
are auxiliary fields. Replacing also these fields by their
equations of motion gives:
\be \label{schematic}
\Delta \gam[\mu,\mub,\phi] = S_L[\phi,\mu,\bar{\mu}]+K[\mu,\bar{\mu}].
\ee
Here
\be \label{lio}
S_L = \deel{k}{4\pi}\int d^2z \,\,
\sqrt{-\hat{g}}\left( \hat{g}^{ab}\dif_a \phi \dif_b \phi + 4e^{-2\phi}
+\phi\hat{R} \right),
\ee
is the well known Liouville action, the metric $\hat{g}$ is defined
by $ds^2=|dz + \mu d\bar{z}|^2$, and $K[\mu,\bar{\mu}]$
reads
\be \label{bkano}
K[\mu,\bar{\mu}] = \deel{k}{4\pi}\int d^2z\,\,
(1-\mu\bar{\mu})^{-1}\left( \dif \mu \dbar \mub -\deel{1}{2}
\mu (\dbar \bar{\mu})^2-\deel{1}{2}\bar{\mu}(\dif \mu)^2\right).
\ee
$\Delta \gam[\mu,\mub,\phi]$ as given in (\ref{schematic}) is known
as the Quillen--Belavin--Knizhnik anomaly \cite{bekn}.
It `covariantizes'
the product of the wave-functions $\Psi[\mu]\,\Psi[\mub]$. In
conclusion we find that Polyakov's result for the covariant
induced gravity action (\ref{polyak}) admits the following
decomposition:
\be \label{decompos}
\frac{c_m}{24\pi}\int R\,\frac{1}{\Box}R=
\Delta \gam[\mu,\mub,\phi]+\gam[\mu]+\gam[\mub].
\ee

In section~4 the above results for ordinary gravity will be
generalized to the case of $W$ gravity, simply by replacing
$Sl(2,\re)$ by some other real, non-compact Lie group.
We have kept our notation quite general so that (\ref{delgamt})
is in fact already the correct formula for general $W$ gravity;
one only has to choose some $G$ and some $\Lambda$ and then the
Fourier transform of (\ref{delgamt}) gives the local counterterm
which covariantizes the product of the chiral wave-function for
general $W$ gravity times its partner of opposite chirality.
Before we end this section we will make a few more comments on
the partition function of covariant induced gravity.

\newsubsection{The Partition Function of Covariant Induced
Gravity}

In the previous subsection we already argued that the partition
for covariant induced gravity could be seen as the norm squared
of the chiral wave-function $\Psi[\mu]=\exp(-\gam[\mu])$,
which is a solution of the chiral Virasoro Ward identity,
when one takes a suitable definition for the inner-product
(\ref{defip}). Equivalently, the partition function could be
written as the modulus squared of the Fourier transformed
wave-function $\Upsilon[T]=\exp(-\gam[T])$, again for some suitable
definition of the inner-product (\ref{defip2}).
In both cases this choice of the inner-product is
such that
\be \label{ofcourse}
{\cal Z}_{grav}=\int \frac{D\mbox{(Fields)}}{\mbox{(gauge vol)}}
\exp(-S_{cov}(\mbox{Fields})),
\ee
where we have to divide by the volume of the diffeomorphism group.
Now looking back at the case where our fields were $T,\bt,G$, we
find that the partition function can be written as:
\be \label{jazokandieook}
{\cal Z}_{grav}=\int \frac{DgDGD\gb}{\mbox{(gauge vol)}}
\delta(J^{+}_z-1)\delta(\bar{J}^{-}_{\bar{z}}-1)
\exp(-kS^{-}_{wzw}(gG\gb^{-1})),
\ee
where $J_z=g^{-1}\dif g$, $\bar{J}=\gb^{-1}\dbar \gb$
and now (gauge vol) denotes the volume of the symmetry group of
the left, right constrained WZW model.
Note that due to the delta functions in (\ref{jazokandieook})
this symmetry group is not simply $G \times G$. What we do have
is invariance under $g\rightarrow gB^{-},\,\,\,G\rightarrow
(B^{-})^{-1}GB^{+},\,\,\,\gb \rightarrow \gb B^{+}$,
where $B^{-}(B^{+})$ is the Borel subgroup of lower (upper)
triangular matrices,
since these transformations do not alter the arguments of
the delta functions.
Using this $B^{-}$ invariance in the left sector and the $B^{+}$
invariance in the right sector we can
bring $g$ in the form such that $g^{-1}\dif g=\Lambda
+ W$, and similarly, bring $\gb$ in the form such that
$\gb^{-1}\dbar \gb=\blam+\bar{W}$. Once this is done
(\ref{jazokandieook}) reduces to\footnote{Again
we neglect a Jacobian in going from $g$ to $T$, which
now equals $\mbox{det}(\dif^3+2T\dif+\dif T)=\exp(-2\gam[T])$. As
before, this contribution can be discarded in the large $k$ limit.}.
\be \label{reduct}
\int \frac{DTDGD\bar{T}}{\mbox{(gauge vol)}}\exp(-S_{cov}(T,\bt,G)).
\ee
The residual gauge invariances
of (\ref{reduct}) are those which leave the form of $g^{-1}\dif g$
and $\gb^{-1}\dbar \gb$ invariant, and as we know from section~2.2
these are just the diffeomorphisms, represented as $g\rightarrow
gX(\eps)$ and $\gb \rightarrow \gb \bar{X}(\beps)$. So altogether
we have shown that (\ref{jazokandieook}) indeed gives the partition
function for covariant gravity.

On the other hand, (\ref{jazokandieook}) can also be trivially
computed to be
\be \label{triviaal}
{\cal Z}_{grav}=\int DG \exp(-kS^{-}_{wzw}(G)) \equiv
{\cal Z}_{wzw}.
\ee
So we come to the remarkable conclusion that, at least in the
large $k$ limit, the partition function for covariant induced
gravity equals the partition function of the $Sl(2,\re)$ WZW
model.

This equivalence could have been anticipated from our previous work
\cite{jj2}, see also \cite{herman}, in which the covariant action
was constructed from \cs. In this theory one works with
wave-functionals $\Psi$ satisfying the Gauss law constraint
$F(A_z,\abar)\Psi=0$. Here $A_z,\abar$ obey the Poisson
brackets: $\{ \abar^{a}(z),A_z^{b}(w) \} = \deel{2\pi
i}{k}\eta^{ab}\delta(z-w)$. The precise form of the wave-functional
$\Psi$ depends on which `polarization' we choose. By this we mean
that we have to divide the set $(A_z^{a},\abar^{a})$ into two
subsets. One subset will contain fields $X_i$ and the other will
consist of derivatives $\vars{}{X_i}$. For instance if one works in
the so called `standard' polarization, in which the $A_z^a$ label the
fields and correspondingly the $\abar^a$ are represented as
derivatives w.r.t. these fields, the solution to the Gauss law
constraint is given by $\Psi[A_z]=\exp(-kS^{-}_{wzw}(g))$, with
$A_z=g^{-1}\dif g$, a result already mentioned in section~2.1.

Furthermore, in
\cs\ there is a natural inner-product for the wave-functional
$\Psi$ given by:
\be \label{ip100}
\left< \Psi | \Psi \right> = \int \frac{DA_zD\abar}{\mbox{(gauge
vol)}} \exp(\act{A_z\abar})\Psi[\abar]\Psi[A_z],
\ee
where we divide by the volume of the gauge group.
Using $\Psi[\abar]=\exp(-kS^{+}_{wzw}(\gb))$, with
$\abar=\gb^{-1}\dbar \gb$, this inner-product can be worked
out as (see also \cite{witten2}):
\be \label{ip200}
\left< \Psi | \Psi \right> = \int DG \exp(-kS^{-}_{wzw}(G))
\equiv {\cal Z}_{wzw},
\ee
where we once more invoked the Polyakov--Wiegman formula, to
change variables from $g,\gb$ to $G=g \gb^{-1}$.
The point is now that instead of the standard polarization
one can choose a different `mixed' polarization in which the
fields are taken from both the $A_z^a$ and the $\abar^a$.
In \cite{jj2} we showed that such a non-standard polarization
can lead to the situation in which the Gauss law constraint for the
wave-functional reduces to the Virasoro Ward identity
(\ref{Virasoro}), and in which the above inner-product (\ref{ip100})
reduces to the inner-product defined in the previous subsection
(\ref{defip}). This then gives the equivalence of the partition
functions for covariant gravity and $Sl(2,\re)$ WZW theory.

\newpage

\newsection{Review of $W$ algebras}

In section~2.2, we saw how imposing constraints
on an $Sl(2,\re)$ connection $A_z$ naturally led to the Virasoro
algebra (\ref{Virasoro}). This algebra basically arose, because
the constraints enabled us to solve some of the zero-curvature
equations (\ref{zerocurv}) explicitly. In this section we will
describe a generalization of this procedure, leading to
$W$ algebras as a generalization of the Virasoro algebra.
The relation between zero-curvature equations and $W$ algebras
has been worked out for the $W_N$ algebras in \cite{bilal1}. A
different approach is to use Hamiltonian reduction to construct $W$
algebras \cite{beroog,adam,ds,hamred}.

The main new result of this section is the introduction of an
operator $L$ that enables us to write down many explicit
expressions, \eg (\ref{eq:abar3}) for $\abar$, and that will
play a prominent role in section~5~ where we discuss global
aspects of $W$ algebras. The $W$ algebras discussed here all
result from embeddings of $sl_2$ in a Lie-algebra $\bg$. We
believe that the reason for the occurrence of these
$sl_2$ algebras is that they describe the identification of
$sl_2$ rotations in the tangent space to the two-dimensional
world-sheet with certain gauge transformations, a picture first
advocated by Polyakov who called this procedure `soldering'
\cite{pol3}. Generic $W$ algebras from $sl_2$ embeddings were
first studied in \cite{adam}.  For a more detailed account of
$W$ algebras, see \cite{wrev}.

Starting with a non-compact semisimple real Lie-algebra
${\bf g}$, we would like to impose some
set of constraints on $A_z$ in order to find an interesting
constrained algebra. However, such constraints will not in
general enable us to solve part of the zero-curvature equations,
or enable us to solve them only at the cost of introducing
infinite power series or
negative powers of components of $A_z$, and this is certainly not
what one wants. To avoid this, one has to choose these
constraints in a special way. To avoid the presence of
denominators containing components of $A_z$ or $\abar$, it would
be nice if the zero-curvature equations that one wants to
solve at each instant were of the form
\ba
&\mbox{{\rm linear combination of certain components of }} \abar  = &
\nonumber \\
& \mbox{{\rm function of }} A_z \mbox{{\rm\ and
the remaining components of }} \abar.  & \label{prototype}
\ea
Clearly, in this case one only would have to solve some linear
set of equations for the components of $\abar$ each time,
thereby avoiding unpleasant denominators. The only way linear
combinations of components of $\abar$ can enter the
zero-curvature equations, is when part of $A_z$ is constant. Let
us therefore try to impose the constraint $A_z=\Lambda+W$, where
$\Lambda$ denotes a constant element of $g_{\bf C}$ and $W$
contains the fields that will generate the constrained algebra
in the end. In this case we can  write (\ref{zerocurv}) in
the form (\ref{prototype}),
\be
-\ad{\Lambda}(\abar)=\dif\abar-\dbar W + [W,\abar].
\label{prototype2}
\ee
The right hand side of this equation will, in general, contain the
same components of $\abar$ as the left hand side, even when we
restrict our attention to only a few components of this
zero-curvature equation. The occurrence of such equations can be
prevented if the Lie algebra comes with a gradation. Recall that
a gradation is a decomposition of the Lie algebra as ${\bf g}=
\bigoplus_{\alpha\in I} {\bf g}_{\alpha}$, where  $I$ is a
finite set of real numbers ($\bg_{\alpha}=\{0\}$ for
$\alpha\not\in I$) and $[{\bf g}_{\alpha},{\bf
g}_{\beta}] \subset {\bf g}_{\alpha+\beta}$. An arbitrary
element $X\in{\bf g}$ can be decomposed as $X=\sum_i X_i$ with
$X_i \in {\bf g}_{\alpha_i}$, $X_i\neq 0$. Let $\maxdeg(X)$
denote $\max_i \alpha_i$ and $\mindeg(X)=\min_i \alpha_i$. If
$\mindeg(\Lambda)>0$ and $\mindeg(\Lambda)>\maxdeg(W)$, then the
left hand side of (\ref{prototype2}) will only contain
components of $\abar$ of lower degree than the components of
$\abar$ occurring on the right hand side of (\ref{prototype2}).
In this case we can express components of $\abar$ in terms of
components of higher degree. Therefore, many of the components
of $\abar$ can be expressed in terms of others if we first
solve the zero-curvature equations of highest degree, and then
take equations of lower and lower degree.
Clearly, the left hand side of (\ref{prototype2}) does not
change if we add an element of $\ker(\ad{\Lambda})$ to $\abar$.
Therefore, once we fix the part of $\abar$ that is annihilated
by $\ad{\Lambda}$, the remaining part of $\abar$ can be expressed
in terms of these independent variables. In the case of the
Virasoro algebra this independent component is $\mu$.

The bilinear product on ${\bf g}$ is given by the Killing form
$(X,Y)=\tr(\ad{X}\ad{Y})=\deel{1}{2h} \mbox{\rm tr}(\ad{X}\ad{Y})$,
normalized such that the longest roots of $\bg$ have length
squared two: $h$ is the dual Coxeter number and $\tr$ represents
the `ordinary' trace.
The Lie algebra ${\bf g}$ has a natural involution, the Cartan
involution, which we will
denote by $\dagger$, and satisfies $(X,X^{\dagger})\geq 0$.
In the case of $sl_N$ this corresponds to
taking the transpose of a matrix in the fundamental
representation. In general, if we decompose the Lie algebra as
$\bg={\bf k}\oplus{\bf p}$, where ${\bf k}$ is the maximal
compact subalgebra of $\bg$, the Cartan involution is $-1$ on
${\bf k}$ and $+1$ on ${\bf p}$.

If we restrict the zero-curvature equation (\ref{prototype2}) to
the complement of $\im(\ad{\Lambda})$ in ${\bf g}$, then the left
hand side of (\ref{prototype2}) vanishes, and the equations that
remain are analogous to (\ref{Virasoro}). These equations are the
Ward identities of the constrained algebra, and at the same time
provide the Poisson brackets. To get a good constrained algebra,
the independent components in $\abar$ must be in one-to-one
correspondence with the components of $W$. The correspondence is
that a gauge transformation with parameter $\abar$ should
generate the same transformations as the `charge'
$Q=\int\,dz(W,\abar)$ generates with respect to the Poisson
bracket. Therefore,
an independent component $m(z,\bar{z})X$ of $\abar$
corresponds to the component $(X,W)$ of $W$. The independent
components of $\abar$ span $\ker(\ad{\Lambda})$, and this
implies that for a proper one-to-one correspondence with
components of $W$ we need
\be \label{eq:propW}
(\ker (\ad{\Lambda})(\bg))^{\bot} \bigoplus \bg_W = \bg, \,\,\,\,\,\,
\dim(\bg_W)=\dim(\ker(\ad{\Lambda})(\bg)),
\ee
where $\bg_W$ is the subspace of $\bg$ spanned by $W$ as we let
its independent components vary.
The zero-curvature equation
(\ref{prototype2}) consists of precisely two pieces: if we
write $F=F_1+F_2$, with $F_1 \in \im(\ad{\Lambda})({\bf g})$ and
$F_2 \in \bg_W$, then $F_1=0$ are
the equations that can be used to express $\abar$ in terms of a
set of independent components, that are in one-to-one
correspondence with $\ker(\ad{\Lambda})$, and $F_2=0$ are the
equations that provide the Poisson brackets of the constrained
algebra. The decomposition of the zero-curvature equations in
two pieces requires $\im(\ad{\Lambda})\oplus\bg_W=\bg$, and
consistency of this equation with (\ref{eq:propW})
gives an extra constraint on $\Lambda$, namely
\be
\label{eq1}
\ker(\ad{\Lambda})({\bf g})^{\dagger} \bigoplus
\im(\ad{\Lambda})({\bf g}) = {\bf g}.
\ee
We expect that if this condition is satisfied, one will get a
good constrained algebra. This algebra does not necessarily
contain the Virasoro algebra. This will only be the case if the
constraints are compatible with conformal invariance, which
will impose extra conditions on the choice of $\Lambda$.
The connection $A_z$ transforms as a spin 1 field under
co-ordinate transformations, whereas a constant like $\Lambda$
transforms as a spin 0 fields. Therefore one cannot just put
$A_z$ equal to $\Lambda+W$. The only way to get things right is
to change the conformal weight of $A_z$ so that $\Lambda$ will
have spin 0. Changing the conformal weight of $A_z$ is in
one-to-one correspondence with adding improvement terms to the
Sugawara energy-momentum tensor of the type $-\tr(H_0 \dif J)$,
where $H_0$ is an arbitrary element of the Cartan subalgebra of
${\bf g}$. These define, in turn, just a gradation of ${\bf g}$,
where the decomposition ${\bf g}=\bigoplus_{\alpha} {\bf
g}_{\alpha}$ is precisely the decomposition of ${\bf g}$ in
eigenspaces of $\ad{H_0}$. Such an improvement term will
therefore change the conformal weight of $X_{\alpha}\in {\bf
g}_{\alpha}$ from 1 to $1-\alpha$. This shows that we need a
gradation of ${\bf g}$ such that $\Lambda$ will be homogeneous
of degree 1. If we write \be
W=\sum_{i=1}^n W^i(z,\bar{z}) \label{weq}
X_i, \ee where $X_i\in\bg_{\alpha_i}$, then $W^i(z,\bar{z})$
will have weight $s_i=1-\alpha_i$. Let us summarize the
conditions we have obtained so far, so as to obtain a proper
constrained algebra:
\begin{itemize}
\setlength{\itemindent}{-5mm}
\item $\Lambda$ is constant and homogeneous of degree one with
respect to some gradation of $\bg$.
\item $\ker(\ad{\Lambda})^{\dagger}\bigoplus
\im(\ad{\Lambda})=\bg$.
\item $\bg_W \bigoplus \im(\ad{\Lambda}) = \bg$, and $\dim(\bg_W\cap
\bg_{\alpha})=
  \dim(\ker(\ad{\Lambda})\cap\bg_{-\alpha})$.
\item $\min\{s_i\}>0$, so that $1=\deg(\Lambda)>\maxdeg(W)$.
\end{itemize}
The third condition guarantees that $W$ admits a decomposition
of the type (\ref{weq}).
At this stage we can already derive a general relation between
the spins $\{s_i\}$ of the constrained algebra, and the
dimension of $\bg$. Let $n_{\alpha}=\dim(\ker(\ad{\Lambda}) \cap
\bg_{\alpha})$, $m_{\alpha}=\dim(\im(\ad{\Lambda}) \cap
\bg_{\alpha})$ and $d_{\alpha}=\dim(\bg_{\alpha})$. The fact that
$\Lambda$ is homogeneous of degree 1 gives the relation
$d_{\alpha}=n_{\alpha}+m_{\alpha+1}$, and
(\ref{eq1}) implies $d_{\alpha}=n_{\alpha}+m_{-\alpha}$. Subtracting
these two identities gives $m_{\alpha+1}=m_{-\alpha}$. An
immediate consequence is that $\sum_{\alpha}(2\alpha+1)m_{-\alpha}=0$,
as $(2\alpha+1)$ is odd under $-\alpha \rightarrow \alpha+1$.
Furthermore $\deg(X)=-\deg(X^{\dagger})$  implies
$d_{\alpha}=d_{-\alpha}$, so that $\sum_{\alpha} \alpha
d_{\alpha}=0$. Combining all this we have
\ba
\sum_i (2s_i-1) & = & \sum_{\alpha} n_{-\alpha}
(1-2\alpha)=\sum_{\alpha} n_{\alpha} (1+2\alpha)=\sum_{\alpha}
(d_{\alpha}+m_{-\alpha})(1+2\alpha) \nonu & = &
\sum_{\alpha} d_{\alpha} =
\dim\,\bg.
\label{spinid}
\ea
This hints at the existence of an underlying $sl_2$ structure, as the
dimension of a spin $(s-1)$ representation is precisely $(2s-1)$.
Equation (\ref{spinid}) would follow trivially if we would have
decomposed $\bg$ with respect to some $sl_2$ subalgebra. Later
we will see how this $sl_2$ structure emerges, once we require
that the constrained algebra contains the Virasoro algebra.

The choice of $W$ is severely restricted by the third of the
four conditions mentioned above. Although the subspace
$\bg_W\subset\bg$ has to fulfill the above criteria,
one can still choose different $\bg_W$. Different
choices of ${\bg_W}$ will not affect the weight spectrum $\{s_i\}$,
but will correspond to basis transformations of the constrained
algebra. Different choices of ${\bg_W}$ are in the literature known as
different Drinfeld-Sokolov gauges. One particular simple choice
is the `highest weight gauge', corresponding to choosing
${\bg_W}=\ker(\ad{\Lambda^{-}})$, for a particular $\Lambda^-$
that we will define later.
In this gauge all the fields
$W^i(z,\bar{z})$ will automatically be primary with respect to
the Virasoro subalgebra. However, in some cases it is
advantageous to choose a different gauge, for instance one in
which the non-linear algebra formed by the $W^i$ is manifestly
quadratic \cite{ds,jj2}. In \cite{jj2} this quadratic form was
necessary to be able to obtain a covariant action for
$W_N$ gravity from \cs.

Given such a choice of subspace ${\bg_W}$, it will turn out to be very
convenient to introduce an operator $L$ which plays the role of
the inverse of $\ad{\Lambda}$. More precisely, let ${\bg_W}^{\bot}$ be
the orthocomplement
of ${\bg_W}$ in $\bg$, then $\ad{\Lambda}$ defines
an invertible linear operator from
${\bg_W}^{\bot} \rightarrow ({\bg_W}^{\bot})^{\dagger}$.
Let $L$ be the
inverse of this operator, extended by $0$ to a linear operator
$\bg\rightarrow \bg$. Let us also introduce a set of projection
operators in the Lie algebra $\bg$, called $\pri,\prk,\prid$ and
$\prkd$, being the orthogonal projections in $\bg$ onto the
subspaces $({\bg_W}^{\bot})^{\dagger},{\bg_W}^{\dagger},
{\bg_W}^{\bot}$ and ${\bg_W}$
respectively. The subscripts $i$ and $k$ refer to image and
kernel, as e.g. $\pri$ is almost the projection onto the image
of $\ad{\Lambda}$. However, as we allow for other than the
highest weight gauges, one should keep in mind that $\pri$ need
not be exactly equivalent to the projection onto the image of
$\ad{\Lambda}$.

Using the operator $L$, it is possible to derive an expression
for $\abar$ where we express it in terms of its independent
components. If we apply $\pri$ to (\ref{prototype2}) we find
\be \label{eq:abar}
-\pri \circ \ad{\Lambda} \abar=\epsilon \dif \pri \abar + \epsilon
\pri [W,\abar],
\ee
where we introduced a `small' parameter $\epsilon$. We will
solve this equation perturbatively in $\epsilon$ and later put
$\epsilon=1$ (and see if that makes sense). To lowest order in
$\epsilon$, the right hand side vanishes and $\abar=F$, where
$F$ is an arbitrary element of $\ker \ad{\Lambda}$. The first order
term of $\abar$ satisfies the equation
\be \label{eq:abar2}
-\pri \circ \ad{\Lambda} \abar^{(1)}=\dif \pri F +
\pri [W,F].
\ee
By definition, $L\circ\pri=L$, and therefore this equation is
solved by $\abar^{(1)}=-L(\dif F + [W,F])$. Proceeding with
higher orders we find in precisely the same way that
$\abar^{(k+1)}=-L(\dif+\ad{W})(\abar^{(k)})$. This shows that
\be \label{eq:abar3}
\abar=\frac{1}{1+\epsilon L (\dif + \ad{W})} F.
\ee
The operator $L(\dif + \ad{W})$ always lowers the degree, as $L$
has degree $-1$, $\dif$ has degree 0, and $W$ consists of elements
of degree less than one only. This implies that this operator is
nilpotent, and that it makes perfect sense to put $\epsilon=1$
in (\ref{eq:abar3}). The independent components of $\abar$ are
given by an arbitrary element $F$ of $\ker \ad{\Lambda}$.

This enables us to extract the Poisson brackets for the
constrained algebra. In the same way as we did in the beginning
of this section for the Virasoro algebra, we replace $\dbar W$
in the zero-curvature equation by $\delta_F W$, and we get
\be \label{eq:vara}
\delta_F W=(\dif+ \ad{\Lambda}+\ad{W})\frac{1}{1+L(\dif+\ad{W})}F.
\ee
Restricting this to a component $W^i(z,\bar{z})$ of $W$, we find
an expression for $\delta_F W^i$, and the Poisson brackets can
then be read of from the identity
\be \label{eq:pb}
\delta_F W^i=\left\{ \int dz \, Q(z),W^i(w) \right\}_{PB},
\ee
where $Q$ satisfies $\delta Q / \delta W=F$.

We still have not answered the question whether or not this
algebra contains a Virasoro algebra. This is most easily studied
in the highest weight gauge, so we take $\bg_W=\ker
\ad{\Lambda^{-}}$, where $\Lambda^-$ is yet to be specified.
A natural candidate for the spin 2
energy-momentum tensor is given by the Sugawara-like expression
\be \label{eq:t}
T=\hf ( \Lambda+W,\Lambda+W)-(\rho,\dif W),
\ee
where $\rho\in\prkd H$, with $H$ the Cartan subalgebra of ${\bf g}$,
represents an arbitrary `improvement' term. To compute the
Poisson bracket of $T$ with an arbitrary field $W^i$, we have to
compute (\ref{eq:vara}) with $F$ chosen in such a way that it
generates small $T$-transformations. In other words, $F$ must
satisfy
\be \label{eq:F}
F=\frac{\delta Q}{\delta W}=\frac{\delta( T(z)\epsilon(z))}{\delta W}=
\Lambda\epsilon+(\prk W)\epsilon+\rho\epsilon',
\ee
where we used that $(W,W)=(\prk W,\prk W)$. From (\ref{eq:F}) we
read off that we must choose
$F=\Lambda\epsilon+(\prk W)\epsilon+\rho\epsilon'$.
Our next task is to compute
(\ref{eq:vara}) for this $F$. It turns out that the calculations
simplify a great deal if we demand that $\{\Lambda,-L\Lambda,
\Lambda^{-}\}$ form a $sl_2$ algebra, \ie\
\ba
[\Lambda,\Lambda^{-}] & = & -L\Lambda, \nonu
[\Lambda,L\Lambda] & = & \Lambda , \nonu
[\Lambda^{-},L\Lambda]& = & -\Lambda^{-} .
\label{eq:sl2}
\ea
We do not know whether it is possible to identify a Virasoro
algebra without making these assumptions. If (\ref{eq:sl2}) are
satisfied, it is clear that we have essentially decomposed $\bg$
in representations of this $sl_2$ algebra, and kept one field
$W^i$ for every representation. In the highest weight gauge, one
precisely keeps the `highest weight' of each representation,
namely the one annihilated by $\Lambda^{-}$. Assuming the
validity of (\ref{eq:sl2}), one can step by step compute
$\abar(F)$, with $F=\Lambda\epsilon+(\prk W)\epsilon+\rho
\epsilon'$: $L(\dif+\ad{W})(F)=(L\Lambda)\epsilon'-(\prid W)
\epsilon$, $(L(\dif+\ad{W}
))^2(F)=(L^2\Lambda)\epsilon''=-\Lambda^{-}\epsilon''$,
and all higher powers of $L(\dif + \ad{W})$ acting on $F$ vanish.
This shows that
\ba
\label{eq:abart}
\abar(F) & =  & \Lambda\epsilon+(\prk W)\epsilon+\rho\epsilon'+
(\prid
W)\epsilon-(L\Lambda)\epsilon'+(L^2\Lambda)\epsilon'' \nonu & =
&
\Lambda\epsilon+W\epsilon+\rho\epsilon'
-(L\Lambda)\epsilon'+(L^2\Lambda)\epsilon''.
\ea
Substituting this in the expression for $\delta_F W$ gives
\be
\label{eq:varW}
\delta_{\epsilon}W=-\Lambda^{-}\epsilon'''+\rho\epsilon''+
(W-[\rho-L\Lambda,W])\epsilon'+W'\epsilon.
\ee
This implies that $T$ indeed generates a Virasoro algebra with
nontrivial central extension, and that all other $W^i$ are
primary with respect to this energy-momentum tensor, with
conformal weight given by $1-\mbox{\rm
eigenvalue}(\rho-L\Lambda)$. An exception form the $W^i$ that
live in the Cartan subalgebra of $\bg$, as they in general will
have a background charge with respect to $T$, determined by the
choice of $\rho$. Only if $\rho=0$ will this background charge
vanishes, and will the $W^i$ that live in the Cartan subalgebra
really have conformal weight one. We also see that the gradation
of the Lie algebra that gives the right conformal weight
assignments, is given by the element $H_0=\rho-L\Lambda$ of the
Cartan subalgebra. Because $[H_0,\Lambda]=\Lambda$, $\Lambda$ is
indeed homogeneous of degree one with respect to this gradation.

As for the central charges of these algebras, they are
fixed to a particular value. Arbitrary central charges can
easily be obtained by imposing precisely the same constraints on
$kA_z$ instead of $A_z$. This is equivalent to imposing the
constraints on the currents of a level-$k$ Kac-Moody algebra.
The new Poisson brackets can be obtained from the ones we have
constructed so far by rescaling $W\rightarrow W/k$,
$T\rightarrow T/k$, $\Lambda\rightarrow \Lambda/k$,
$\Lambda^-\rightarrow k\Lambda^-$ and $L\rightarrow kL$. After
rescaling (\ref{eq:t}) and (\ref{eq:varW}) one can easily
extract the central charge for the $W$ algebra, it is given by
\be \label{def:c}
c=-12 (\Lambda,\Lambda^-) k.
\ee

Note that to obtain the usual $W_N$ algebra, one starts
with $\bg=SL_N$ and usually takes
\be \label{eq:l}
\Lambda = \left( \begin{array}{cccccc} 0 & 1 & 0 & \cdots & 0 & 0
\\ 0 & 0 & 1 & \cdots & 0 & 0 \\ \vdots & \vdots & \vdots & & \vdots
& \vdots \\ 0 & 0 & 0 & \cdots & 1 & 0
\\ 0 & 0 & 0 & \cdots & 0 & 1 \\ 0 & 0 & 0 & \cdots &
0 & 0 \end{array} \right),
\ee
which can be compactly notated as $\Lambda=\sum_{i=1}^{N-1}e_{i,i+1}$,
where $e_{i,j}$ is the matrix which has a one on its $i^{th}$
row and its $j^{th}$ column and is zero everywhere else.
An easy computation shows that
(\ref{eq:sl2}) is satisfied, if $\Lambda^{-}=\hf \sum_{i=1}^{N-1}
i(N-i)e_{i+1,i}$.
The corresponding $sl_2$-embedding is a special case of a
`principal' $sl_2$ embedding \cite{kost}. Plugging this
$\Lambda$ and $\Lambda^-$ back into (\ref{def:c}) gives
$c=-k(N^3-N)$.

So far we have imposed several constraints in order that the
constrained algebra will have certain desired properties. If we
omit some of these constraints, one may still end up with a good
algebra, although it will not necessarily have all the
properties described above. For instance, if we do not require
the existence of an $sl_2$ subalgebra, the constrained algebra
is still ok, but it need not have a Virasoro subalgebra.
In other cases one can impose new constraints on the
$W$ algebras and reduce them even further, as \eg\ in
\cite{wlnpol}. We will in this paper not concern us with such
types of algebras, and focus our attention only on those that
can be obtained from a $sl_2$ embedding, although some of the
results may also be applicable to more exotic cases. On the
other hand, if a $W$ algebra contains the Virasoro algebra it
automatically has some kind of $sl_2$
structure \cite{watts}, so that this restriction
does not seem to be very severe.

\newsection{The Action of Chiral and Covariant $W$ Gravity}

In the previous section we explained how general (classical)
$W$ algebras can be obtained by considering constrained gauge
fields $A_z=\Lambda+W$, where $\Lambda$ and $W$ obey certain
criteria. Given a proper choice for $\bg_W=\prkd \bg$,
we showed that the solution $\abar$ of
the zero-curvature condition $F(A_z,\abar)=0$ is given by
\be \label{nogeens}
\abar=\frac{1}{1+L(\dif + \ad{W})}F(\mu_i),
\ee
where $F(\mu_i)$ is an element of the kernel of $\ad{\Lambda}$.
Here the $\mu_i$ are the parameters of the $W$ algebra dual to the
fields $W_i$. They are the generalization of the Beltrami differential
$\mu$ which appeared for the case of ordinary gravity.
To be more precise, the zero-curvature condition for the constrained
gauge field $A_z$ splits into two parts: if we write $F=F_1+F_2$, with
$F_1 \in \pri \bg$ and $F_2 \in \prkd \bg$, then $F_1=0$ are the
equations leading to (\ref{nogeens}) and $F_2=0$ are the equations
which encode the generalized $W$ transformations (\ref{eq:vara}):
\be \label{nogeenx}
\delta_{\eps}W=D_{A_z}(X(\eps)),
\ee
where  $X(\eps)$ is (\ref{nogeens}) with the $\mu_i$ in $F$
replaced by $\eps_i$.
In this section we will consider the chiral and covariant action
for these generalized $W$ algebras.

\newsubsection{The Chiral Action for General $W$ Gravity}

In the spirit of section~2 the induced action for general $W$
gravity can be defined as:
\be \label{spirit}
\exp(-\gam[\mu_i])=\left< \exp -\deel{1}{2\pi}\int d^2z\,
\sum_i \mu_iW_i^{mat} \right>,
\ee
where the $W^{mat}_i$ satisfy the operator product expansions
of the generalized $W$ algebra. These operator product
expansions give differential equations for the chiral induced
action $\gam[\mu_i]$, as we explained in detail in section~2.
Since these differential equations can be obtained
from the zero-curvature condition for a constrained gauge field,
and since the solution to this zero-curvature
condition is given by the WZW functional, we have as before that
$\gam[W_i]$, the Fourier transformed of $\gam[\mu_i]$, is given
by $\gam[W_i]=-kS_{wzw}^{-}(g)$, where $g$ is restricted such
that $g^{-1}\dif g=\Lambda+W$, and $k$ is related to the matter
central charge via $c_m \sim -12(\Lambda,\Lambda^-)k$ as shown in
section~3.
Repeating calculation (\ref{ft})
of section~2.3, we compute $\gam[\mu_i]$ from this by Fourier
transformation:
\ba \label{encore}
\exp(-\gam[\mu_i]) & = & \int DW_i
\exp(-\gam[W_i]-\act{WF(\mu_i)})
\nonu
& = &
\int Dh \,\delta\! \left(\prk (F(\mu_i) - f\jbar f^{-1}) \right)
\exp(-kS_{wzw}^{+}(h)),
\ea
where $\jbar=h^{-1}\dbar h$ and $f=\exp(-z\Lambda)$.
We thus find that the correlation functions of chiral $W$ gravity
can be obtained from WZW theory, if one identifies the $\mu_i$
with Kac-Moody currents in the way dictated by the delta-function in
(\ref{encore}).

Taking the large $k$ limit of (\ref{encore})
we find (using the results of
Appendix~A) that $\gam[\mu_i]=kS^{+}_{wzw}(hf)$,
where $h$ is such that $h^{-1}\dbar h$ is given
by the r.h.s. of (\ref{nogeens}). So the equation
of motion for $h$ is such that the pair $h^{-1}\dbar h$ and
$A_z=\Lambda+W$ have zero-curvature.

\newsubsection{The Covariant Action for General $W$ Gravity}

Recall from section~2.5 that in order to find the covariant
action for general $W$ gravity, we have to construct a local
counterterm whose anomalous behavior cancels that of the chiral
actions. As in the case of ordinary gravity this local
counterterm is most easily found if we work with the Fourier
transformed actions $\gam[W_i]$ instead of the $\gam[\mu_i]$.
In fact, (\ref{delgamt}), derived in section~2.5 for the
case of ordinary gravity, is already the correct
formula for the local counterterm for general $W$ gravity,
once we take the appropriate choice for $\Lambda$,
$W$ and $G$. So the local counterterm is given by:
\ba \label{delgamW}
\Delta \gam[W_i,\bar{W}^i,G] &=&
\act{(\blam+\bar{W})G^{-1}\dif G}
-\act{(\Lambda+W)\dbar G G^{-1}} \nonu
& &+\act{(\Lambda+W)G(\blam+\bar{W})G^{-1}}-kS^{-}_{wzw}(G).
\ea
Notice the close resemblance of this expression with the action
for a gauged WZW theory \cite{gwzw}. In fact, apart from a term
$\act{(\Lambda+W)(\blam+\bar{W})}$, it is precisely the action
for a gauged WZW theory, where the gauge fields are of a
particular restricted form, such that one does not gauge a
subgroup, but one actually gauges the $W$ algebra. An important
difference, however, is that the covariant action
is invariant under both left and right $W$ transformations,
whereas the gauged WZW model only is invariant under left and
right gauge transformations simultaneously.

The local counterterm $\Delta \gam[\mu_i,\mub_i,G]$ can now be
obtained from (\ref{delgamW}) by Fourier transformation with
respect to $\mu_i,\mub_i$. As said before this Fourier
transformation is simple and exact to do, since the $W$ fields
appear algebraically and at most quadratic in (\ref{delgamW}).
One easily verifies that the saddle-point equations for the
$W$ fields are:
\ba \label{zadel}
\prk  \left( \dbar G G^{-1}-G(\blam+\bar{W})G^{-1}+F(\mu_i)
\right) &=& 0, \nonu
\prkd \left( G^{-1}\dif G +G^{-1}(\Lambda+W)G-\bar{F}(\mub_i)
\right) &=& 0.
\ea
Unfortunately, there is no general formula for the solutions of these
saddle-point equations which is valid for all generalized $W$ algebras.
We can give a general formula for the case of the standard $W_N$
algebras, valid for all $N$, but
for the non-standard $W$ algebras the solutions to (\ref{zadel})
should be determined for each case separately.

For standard $W_N$ we take $Sl(N,\re)$ as our gauge group,
and we constrain the gauge field as follows \cite{ds}:
\be \label{wnconstr}
A_z = \Lambda+W=\left( \begin{array}{cccccc} 0 & 1 & 0 & \cdots & 0 & 0
\\ 0 & 0 & 1 & \cdots & 0 & 0 \\ \vdots & \vdots & \vdots & & \vdots
& \vdots \\ 0 & 0 & 0 & \cdots & 1 & 0
\\ 0 & 0 & 0 & \cdots & 0 & 1 \\ W_N & W_{N-1} & W_{N-2} & \cdots &
W_2 & 0 \end{array} \right).
\ee
Note that $\prk \bg$ is spanned by $\{ e_{i,N} \}$, for
$i=1,\ldots,N-1$, where $e_{i,j}$ is the matrix
which has a $1$ on its $i^{th}$ row and its $j^{th}$ column and
is zero everywhere else. $F(\mu_i)$ can be conveniently
parametrized by $F(\mu_i)=\sum_{i=2}^N \mu_i \Lambda^{i-1}$.
In this case $G$ admits a decomposition in terms of three
subgroups:
$G=\prkd G \pry G \prk G$, which in terms of matrices looks like
\be \label{Gontb}
G=\left( \begin{array}{cccc} 1 &  &  &  \\  & \ddots &  &  \\
& & 1 & \\ {*} & \cdots & * & 1 \\  \end{array} \right)
\left( \begin{array}{cccc} * & \cdots & * & 0 \\ \vdots &  & \vdots
& \vdots \\ {*} & \cdots & * & 0 \\ 0 & \cdots & 0 & * \end{array} \right)
\left( \begin{array}{cccc} 1 & &  & * \\  & \ddots &  & \vdots \\
&  & 1 & * \\  &  &  & 1 \end{array} \right),
\ee
and which is crucial to be able to solve the saddle-point
equations (\ref{zadel}). One finds
\ba \label{oplzadel}
W &=& (\pry G)\,\prkd \left( \bar{F}(\mub_i)
-\Lambda^G \right)\,(\pry G)^{-1}, \nonu
\bar{W} &=& (\pry G)^{-1}\, \prk \left( F(\mu_i)
-\blam^G\right) \,(\pry G),
\ea
where $\Lambda^G=G^{-1}\Lambda G+G^{-1}\dif G$ and a similar definition
for $\blam^G$.
Substituting this solution for the $W$ fields back into the local
counterterm gives:
\ba \label{cvac}
\Delta \gam[\mu_i,\mub_i,G]  \!\!\! &=& \!\!\!
\act{\Lambda G\blam G^{-1}}+
\act{\blam G^{-1}\dif G}-\act{\Lambda\dbar G G^{-1}} \nonu
&-&\!\!\!\act{\prk(F(\mu_i)-\blam^G)\pry G \prkd(\bar{F}(\mub_i)
-\Lambda^G)\pry G^{-1}}-kS^{-}_{wzw}(G). \nonu
& &
\ea
The total covariant action is given by the sum of this
local counterterm and the chiral action constructed in the previous
subsection together with its partner of opposite chirality. This
covariant action is invariant under the transformations \cite{jj2}
\ba \label{invtransf}
\delta_{\eps}G &=& -X(\eps)G, \nonu
\delta_{\eps}\prk F(\mu_i) &=& \prk \left( \dbar X(\eps)+\lha
\abar,X(\eps)\rha \right) ,
\ea
where $X(\eps)$ is still given by (\ref{nogeens}), but with the $W$
fields in (\ref{nogeens}) replaced by the solutions to
the saddle-point equations (\ref{oplzadel}), and $\abar$ is
(\ref{nogeens}) with the $W$ fields given by
$W_i=\deel{2\pi}{k}\vars{\gam[\mu_i]}{\mu_i}$.
These transformations form a $W_N$ algebra by construction.

\newsection{Global Aspects of $W$ Algebras}

So far we have restricted our attention to $W$ algebras and
actions that were defined on the complex plane. If we want to
answer questions like `what is the moduli-space of
$W$ algebras', we have to discuss $W$ algebras on higher genus
Riemann surfaces, as on the complex plane one can choose
globally well-defined co-ordinates, and it is sufficient to
express everything in terms of these co-ordinates. One need not
bother about the transformation properties of the different
objects one encounters under change of co-ordinates, and
generally there are no moduli. In this section we will make a
start of the study of $W$ algebras on general Riemann surfaces.
In previous sections we have seen how one can use the
formulation of $W$ algebras in terms of zero-curvature equations
and special gauge-transformations to obtain the chiral and
covariant actions for $W$ gravity, and we will now also use the
same formulation to go to arbitrary Riemann surfaces.

Suppose that we have some $G$-bundle $P$ over a Riemann surface
$\Sigma$, equipped with a connection that in certain local
complex co-ordinates can be written as $A=A_z dz+\abar
d\bar{z}$. We want, as before, to impose constraints on $A_z$,
and in particular to set certain components of $A_z$ equal to
some constant value. However, as $A_z$ transforms as a one-form
under co-ordinate transformations, and not as a function, one
cannot impose these constraints globally. The same problem
already exists on the algebraic level, where one wants to put a
spin one current equal to a constant, which is not compatible
with conformal invariance. In the latter case, the problem is
resolved by improving the energy momentum tensor, which
changes the spins of the currents. Luckily this `soldering'
procedure has a geometrical counterpart, which amounts to
twisting a trivial $G$-bundle into a non-trivial one. For the
sake of simplicity, we will restrict our attention to the case
$G=Sl(N,\ce)$ in this section, and assume that we are working in
a highest weight gauge and that the there is no background
charge $\rho$ in (\ref{eq:t}). Let us first describe
how the twisting works.

\newsubsection{The Twisted Bundle}

Improving the energy-momentum tensor with a term $-\tr(H_0 \dif
J)$ amounts to changing the gradation of $\bg$ from the trivial
one into the one determined by $H_0$. Suppose that $H_0=\diag
(d_1,\ldots,d_N)$ in the fundamental representation of $sl_N$,
so that the degree of $E_{\alpha_i}$ is $d_{i+1}-d_i$ for each
simple root $\alpha_i$. Let $K$ be the holomorphic cotangent
bundle $T^*_{\ce}\Sigma$ of the Riemann surface, and
let $E_{ij}$ denote the line bundle $K^{-d_i}$, $i,j=1\ldots
N$\footnote{for noninteger $d_i$ one has to choose some
appropriate root of $K$, the precise choice is not important here.}.
Given (locally) a section $s$ of the $N^2$-dimensional vector
bundle $E=\bigoplus_{i,j=1}^N E_{ij}$, one can consider the
function $\det(s_{ij}):\Sigma\rightarrow \ce$, which is well
defined because $\sum d_i=0$. The sections $s$ with $\det(s)=1$
are sections of a principal $Sl(N,\ce)$ bundle $P_c$, with
reduced structure group $GL(1,\ce)$.
This is precisely the twisted bundle we need.
To see this, consider the adjoint bundle $\add (P_c)$. This bundle
has an alternative description as $sl(V)$, the bundle of
traceless endomorphisms of $V$, sometimes also denoted by $\End_0(V)$,
where $V$ is the vector bundle
\be \label{def:V}
V=K^{-d_1}\bigoplus\ldots\bigoplus K^{-d_N}.
\ee
A connection on $P_c$ is locally a one-form with values in
$\add (P_c)$. The $dz$ part of such a connection looks like
$a_{ij}dz$, where $a_{ij}$ is an $N\times N$ matrix; $a_{ij}$
transforms as a section of
$K\otimes\hom(K^{-d_j},K^{-d_i})\simeq K^{1+d_j-d_i}$. The
improvement term $-\tr (H_0\dif J)$ changes the spin of
$a_{ij}$ from 1 to $1+d_j-d_i$, and we see that by twisting the
principal $Sl(N,\ce)$ bundle $P_c$ we have achieved the same
thing. However, a connection is only locally an $\add (P_c)$-valued
one-form, and globally $a_{ij}$ does not transform as a section
of $K^{1+d_j-d_i}$. Only the difference between two connections
transforms globally in the proper way, because such a
difference is a global $\add (P_c)$-valued one-form\footnote{The
space of connections is an affine space modeled on the vector
space $\Omega^1(\Sigma ; \add (P_c))$.}. This shows that we should
impose constraints on the difference of two connections, rather
than on the connection itself. Before discussing the form of
such constraints, we will now first consider WZW actions for
connections on non-trivial bundles, which are necessary to
write down generalizations of the actions we have considered so
far.

\newsubsection{Generalized WZW Action}

In section~2 we saw that the WZW action arises as the induced
action for a gauge field coupled to some matter system.
In this section we will take a specific matter system
system, namely we will study a gauge field coupled to a chiral
fermion $\psi$. This $\psi$
transform as a section of $V\otimes
(\bar{K})^{\hf}$, and
$A_z$ is part of a connection of $P_c$. The action
\be \label{actagain}
S(\psi,A_z)=\deel{1}{\pi}
\int d^2z\,\tr\left( \psi^{\dagger}
(\dif + A_z) \psi \right)
\ee
is gauge invariant under $\psi^{\dagger}\rightarrow \psi^{\dagger}
h$, $\psi \rightarrow  h^{-1} \psi$, and
$A\rightarrow h^{-1}A h + h^{-1}\dif h$. The two-point function
$G(A_z;z,w)=\langle \psi(z) \psi^{\dagger}(w)\rangle$
satisfies $(\dif+A_z(z))G(A_z;z,w)=\pi\delta(z-w)$. From this
one finds the following rule for the change of $G$ under a gauge
transformation $h$:
$G(A_z^h;z,w)=h(z)^{-1}G(A_z;z,w)h(w)$. For
$z\rightarrow w$ $G$ behaves as
\be \label{asympt}
G\sim \frac{1}{\bar{z}-\bar{w}}+\chi_{\bar{z}}(z)
-A_z(z)\frac{z-w}{\bar{z}-
\bar{w}}+ \mbox{\rm terms that vanish as }
z \rightarrow w.
\ee
Under the gauge transformation $A_z\rightarrow
A_z^h=h^{-1}\dif h + h^{-1}A_z h$ we find, by expanding
$h(w)=h(z)+(\bar{z}-\bar{w})\dbar h(z)+(z-w)\dif h(z)
+\ldots$, that
$\chi_{\bar{z}}\rightarrow h^{-1}\chi_{\bar{z}} h + h^{-1}\dbar h$,
\ie $\chi_{\bar{z}}$
transforms as a connection. Furthermore, if locally
$A_z=0$ we know that $G$ is exactly given by
$1/(\bar{z}-\bar{w})$. Combining these facts we deduce that the
curvature of the connection one-form $A_z dz+\chi_{\bar{z}}
d\bar{z}$ must
vanish. To express this fact in terms of the current
$\jbar$, we must first define what we mean by
$\jbar$. The naive definition
$\jbar=\lim_{z\rightarrow w} G(A_z;z,w)$ does not work,
due to the singularity of $G$ as $z\rightarrow w$. This means we
have to regularize $\jbar$, and a standard way of doing
this is by using point splitting regularization: one defines
$\jbar=\lim_{z\rightarrow w} (G(A_z;z,w)-G_0(A_z;z,w))$
where $G_0$ is some function that has the same singular behavior
as $G$. This means that
\be \label{asympt2}
G_0\sim \frac{1}{\bar{z}-\bar{w}}+B_{\bar{z}}^0(z)
-A_z(z)\frac{z-w}{\bar{z}-
\bar{w}}+ \mbox{\rm terms that vanish as }
z \rightarrow w.
\ee
Here, $B_{\bar{z}}^0(z)$ is some fixed field that transforms as
the $d\bar{z}$-part of a connection. Unfortunately, one cannot just
take $B_{\bar{z}}^0$ equal to zero, because zero is not a
globally well-defined connection. The current $\jbar$ is
\be \label{defcur}
\jbar=\chi_{\bar{z}}-B_{\bar{z}}^0,
\ee
and the Ward-identity (expressing the fact that the current $\jbar$
is not conserved) reads
\be \label{curnoncon2}
D_{A_z} \jbar = \dbar A_z + D_{A_z} B_{\bar{z}}^0.
\ee

This identity should come from some kind of generalized
WZW action. The WZW action $S_{wzw}(g)$ (\ref{deff}) is really a
functional on the space of gauge transformations, and on the
complex plane this space is isomorphic to maps from the plane to
$G$. The terms $g^{-1}\dif g$ in the WZW action are just
$\bn^g$, where $\bn$ is the trivial connection on the complex
plane, and the superscript $g$ refers as usual to a gauge
transformation. Even on the complex plane one sees that the
WZW action is not invariant if one chooses a different
trivialization of the (trivial) bundle $P_c$. If one chooses a
different trivialization, related to the first one via a gauge
transformation $h$, then $\bn^g \rightarrow (\bn^h)^{h^{-1}gh}$,
because $g$ transforms as a section of the adjoint bundle
$\aad(P_c)$\footnote{$\aad(P_c)$ is defined as $(P_c\times G)/G$,
where one the $G$-action on $P_c\times G$ is given by the
standard left action on $P_c$ and by the adjoint action on $G$.}.
However, $(\bn^h)^{h^{-1}gh}=(gh)^{-1}\dif(gh)$, and as is well
known, $S_{wzw}(gh)\neq S_{wzw}(g)$. As the WZW action should
not depend on the choice of trivialization of $P_c$, there is
something wrong with the identification of $g^{-1}\dif g $ with
$\bn^g$. Actually, there is another possibility, which turns out
to be the right one, namely to identify $g^{-1}\dif g$ with
$\bn^g-\bn$. Under a change of trivialization
\be \label{identif}
\bn^g-\bn \rightarrow (\bn^h)^{h^{-1}gh}-(\bn^h)=h^{-1}(\bn^g-
\bn)h,
\ee
and the WZW action is invariant, because the $h$ and $h^{-1}$
cancel each other inside the traces in the WZW action.

If $P_c$ is a non-trivial bundle, one cannot take $\bn$ as a
well-defined connection, and it must be replaced by some other,
fixed connection $B=B^0_z dz + B^0_{\bar{z}} d\bar{z}$. It
turns out that we must require $B$ to be flat. As $A_z$ is
identified with $g^{-1}\dif g$, we must replace $g^{-1}\dbar g$
by a function $\abar(A_z)$ determined by requiring the
curvature of the connection one-form $A(A_z)=A_z dz + \abar
d\bar{z}$ to vanish. This leads to the following definition of
the WZW action
\be \label{wzw2}
k S^{-}_{wzw}(A;B)=\deel{k}{8\pi}\int_{\Sigma}
\,\,\tr((A-B^0)\wedge \ast
(A-B^0)) - \deel{k}{12\pi} \int_M \tr(\tilde{A}-\tilde{B}^0)^3,
\ee
where $\dif M=\Sigma$, and $\tilde{A},\tilde{B}^0$ denote
flat extensions of $A,B^0$ on a bundle $\tilde{P_c}$ on $B$ that
restricts to $P_c$ on $\Sigma$. The $\ast$ is the Hodge star on
the Riemann surface $\Sigma$, where we assume that some metric
compatible with the complex structure on $\Sigma$ is given. That
an extension $\tilde{P_c}$ of $P_c$ exists can be seen as follows
\cite{botttu}: it is sufficient to construct a complex
vector bundle $\tilde{V}$ over $B$, that extends $V$
(\ref{def:V}). The vector bundle $V$ is the pull-back of a
universal vector bundle over a certain grassmannian $Gr$.
The map we
use to pull back this universal vector bundle maps $\Sigma$ into
an element of the second homology of the grassmannian, and it is
precisely this element of $H_2(Gr)$ that gives the obstruction
to construct an extension $\tilde{V}$. Because this element is
essentially the first Chern class of $V$, and the first Chern
class of $V$ vanishes, we know $\Sigma$ maps to zero in
$H_2(Gr)$ and an extension $\tilde{V}$ indeed exists.
It may seem surprising that
one needs an additional connection $B$ to write down the
WZW action, but this is necessary if one wants to write down an
action for chiral fermions only. It also appears when
considering determinant bundles associated to operators such as
$D_{A_z}$ \cite{quillen}.

Let us demonstrate that (\ref{wzw2}) indeed satisfies the
Ward-identity (\ref{curnoncon2}), if we identify the two
$B^0_{\bar{z}}$'s with each other. Consider a small variation
$A\rightarrow A+\delta A$. From the zero-curvature equation
$dA+A\wedge A=0$ we find $d\delta A + A\wedge\delta A + \delta A
\wedge A=0$, with similar equations for $\tilde{A}$. Using this
we compute
\ba
\delta \tr (\tilde{A}-\tilde{B}^0)^3 & = & 3 \tr ( \delta \tilde{A}
\wedge (\tilde{A}-\tilde{B}^0) \wedge (\tilde{A}-\tilde{B}^0) )
\nonu
& = & 3 \tr(\tilde{A}\wedge \delta \tilde{A} \wedge
(\tilde{A}-\tilde{B}^0)  +
\delta \tilde{A}\wedge  \tilde{A} \wedge
(\tilde{A}-\tilde{B}^0)  - \nonu & &
\delta \tilde{A} \wedge \tilde{A} \wedge \tilde{A} +
\delta \tilde{B}^0 \wedge \tilde{B}^0 \wedge \tilde{B}^0 ) \nonu
& = & 3 \tr(-d(\delta \tilde{A})\wedge (\tilde{A}-\tilde{B}^0) +
\delta\tilde{A}\wedge (d\tilde{A}-d\tilde{B}^0) ) \nonu
& = & -3\tr(d(\delta\tilde{A}\wedge(\tilde{A}-\tilde{B}^0))).
\label{varterm2}
\ea
This gives for the total variation of the WZW action
\be \label{varwzw2}
k \delta S^-_{wzw}(A_z;B^0)=
\deel{k}{4\pi}\int_{\Sigma}
\tr(\delta A\wedge (1+\ast)(A-B^0))=
\deel{k}{2\pi}\int_{\Sigma} d^2 z \,\,
\tr(\delta A_z (A-B^0)_{\bar{z}}),
\ee
because $(1\pm\ast)/2$ are precisely the operators that define the
complex structure. This shows that this action indeed solves the
Ward-identity (\ref{curnoncon2}) with $\jbar=2\pi\vars{S^-_{wzw}
(A_z)}{A_z}$, if we identify the $\bar{B}^0$'s with each other.

Having defined a generalized
WZW action, it is interesting to see whether
this action shares some of the properties of the ordinary
WZW action. Using a calculation similar as (\ref{varterm2}), one
can verify the following version of the Polyakov-Wiegmann
formula
\be \label{pw2}
k S^-_{wzw}(A;B)=k S^-_{wzw}(A;C)+k S^-_{wzw}(C;B)+
\deel{k}{2\pi}\int_{\Sigma} d^2 z \,
\tr((A-C)_z(C-B)_{\bar{z}}),
\ee
from which the usual Polyakov-Wiegmann formula follows by
putting $A=(gh)^{-1}d(gh)$, $C=h^{-1}dh$ and $B=0$ for a trivial
bundle $P_c$. Another issue is whether $S_{wzw}(A;B)$ depends on
the choice of extension $\tilde{A}$ and $\tilde{B}$. Choosing a
different extension will change the action by a term $\deel{k}{12\pi}
\int_M \tr(\tilde{A}-\tilde{B})^3$, where now $\Sigma\subset B$ and
$\dif B=\emptyset$. Let ${\cal U}$ denote the space of flat
connections $\tilde{A}$ on such a $B$ such that
$\tilde{A}\mid_{\Sigma}=A$, and consider the function
$r_{\tilde{B}}:{\cal U} \rightarrow \ce$ given by
$r_{\tilde{B}}(\tilde{A})=\deel{k}{12\pi} \int \tr
(\tilde{A}-\tilde{B})^3$. From the identity
\be \label{eq:triple}
\tr((\tilde{A}-\tilde{B})^3+(\tilde{B}-\tilde{C})^3+
(\tilde{C}-\tilde{A})^3)=3\tr(d((\tilde{A}-\tilde{B})\wedge
(\tilde{B}-\tilde{C})))
\ee
it follows that
$r_{\tilde{B}}(\tilde{A})+r_{\tilde{C}}(\tilde{B})=
r_{\tilde{C}}(\tilde{A})$. This implies that
$r_{\tilde{B}}(\tilde{A}+\delta\tilde{A})-
r_{\tilde{B}}(\tilde{A})$ is of third order in
$\delta\tilde{A}$, and therefore that $r_{\tilde{B}}$ is locally
constant on ${\cal U}$, \ie\  $r_{\tilde{B}}$ descends to a
map $\pi_0({\cal U})\rightarrow\ce$. We see that the WZW action
is invariant under a continuous change of the choice of
extension. To find out whether or not $k$ is quantized is not
very easy as it requires knowledge of $\pi_0({\cal U})$.
However, in the case that ${\cal U}/{\cal G}$ is
connected, where ${\cal G}$ is the space of gauge
transformations acting on ${\cal U}$, one can say a little bit
more, using the fact that is this case all connected components
of ${\cal U}$ can be reached from a fixed one using gauge
transformations. To do this, one has to take a slightly different
look at the function $r_{\tilde{B}}$. For any group $G$ one can
write down an element of $H^3(G)$ by extending the
three-form $\omega(X,Y,Z)=\deel{k}{12\pi}\tr(X[Y,Z])$
on the Lie algebra of $G$ all over the group $G$. One can choose
$k$ such that $\omega$ defines actually an (possibly trivial)
element of $H^3(G,\ze)$. This three-form is invariant under
the adjoint action of $G$, and therefore defines an element of
$\tilde{\omega}\in H^3(\aad(\tilde{P_c}),\ze)$ which restricts to
$\omega$ on each fiber. A simple computation now shows that
$r_{\tilde{B}}(\tilde{B}^g)=\int_{M} g^{\ast}\tilde{\omega}$,
which is an integer, because $g^{\ast}\tilde{\omega}$ is an
element of integral cohomology and evaluating such an element on
a three manifold without boundary always gives an integer.
We conclude that
$k$ must sometimes be restricted to those values for which
$\tilde{\omega}$ is an element of integral cohomology
(so that upon quantizing the model
everything is independent of the choice of extension),
but if for instance
$\tilde{\omega}=0$ in $H^3(\aad(\tilde{P_c}),\ze)$ for a $k\neq
0$, $k$ can be taken arbitrarily.

\newsubsection{$W$ Algebras on arbitrary $\Sigma$}

In section~3 we saw how $W$ algebras can be constructed by
imposing the constraint $A_z=\Lambda+W$ on the connection
one-form $A_z$. On a general surface these constraints can only
be imposed locally; when relating different co-ordinate patches,
the connection one-form transforms as $A_z\rightarrow \frac{dz}{dz'}
(h^{-1} A_{z'} h +h^{-1}\dif_{z'} h)$, which does not
preserve the constraint $A_z=
\Lambda+W$. However, due to the special structure of the bundles
for which $A_z$ is a connection, one can always choose a
trivialization such that $h$ is a diagonal matrix. For
these gauge transformations, the special form of $A_z$ is
preserved up to the term $\frac{dz }{dz'}h^{-1}\dif_{z'} h$.
We can get rid of this
term by imposing the constraints on $A_z-B_z$ instead of $A_z$,
where $B_z$ is some fixed connection, in the same spirit as we
did in the previous paragraphs. Now the constraints are
preserved: $\Lambda$ was homogeneous of degree one with respect
to the gradation of $\bg$ and does not transform when going from
one co-ordinate patch to another, while the components of
$W$ transform as fields of certain spins that are also
determined by the gradation. A natural choice for $B$ is the
connection $\nabla=\nabla_z+\bnab$ that comes
directly from the Levi-Civita connection
associated to a fixed metric on the Riemann surface. Let us
consider what happens in this case, and impose the constraint
\be \label{eq:constr}
D_{A_z}=\nabla_z+\Lambda+W.
\ee
We can now repeat the same steps as those that lead
to (\ref{eq:abar3}). An important difference is that $\nabla_z$
and $\bnab$ need not commute with each other, and the
answer contains the curvature
$R_{z\bar{z}}=[\nabla_z,\bnab]$. Besides this,
nothing new happens and one finds that
\be \label{eq:abar4}
\bar{D}_{\abar}=\bnab+\frac{1}{1+ L (\nabla_z
+ \ad{W})} (F-L(R_{z\bar{z}})).
\ee
Since $\nabla_z$ has degree zero, $L(\nabla_z+\ad{W})$ is
nilpotent and this expression is well defined.
It is instructive to work out the zero-curvature equation for
the case where $G=Sl(2,\re)$, and to impose the same constraints as
in section~2 (\ref{restr}). In this case the
vector bundle $V$ in (\ref{def:V}) is $V=K^{-\hf}\bigoplus K^{+\hf}$.
Working in isothermal co-ordinates
where $ds^2=\rho dz d\bar{z}$, one finds that
\be \label{eq:hol}
D_{A_z}=\dif+\add\mats{\hf\dif\log\rho}{1}{T}{-\hf\dif\log\rho},
\ee
and from (\ref{eq:abar4}) that
\be \label{eq:ahol}
\bar{D}_{\abar}=\dbar+\add\mats{\hf\dif\mu+\hf\mu\dif\log\rho}{
\mu}{\mu T - \hf\dif^2\mu+\hf\dif\dbar\log\rho-
\hf\dif(\mu\dif\log\rho)}{-\hf\mu\dif\log\rho-\hf\dif\mu}.
\ee
The remaining zero-curvature equation, which is the
generalization of (\ref{Virasoro}), reads
\be \label{Virasoro2}
(\dbar-\mu\dif-2(\dif\mu))T=
-\hf(\dif-\dif\log\rho)\dif(\dif+
\dif\log\rho)\mu+\hf\dbar(\dif^2 \log\rho-\hf(\dif\log\rho)^2),
\ee
which can be rewritten as
\be \label{Virasoro3}
(\dbar-\mu\dif-2(\dif\mu))T=-\hf(\dif^3+2{\cal R}\dif+(\dif{
\cal R}))\mu+\hf\dbar{\cal R},
\ee
where we introduced the projective connection ${\cal R}=\dif^2
\log\rho-\hf(\dif\log\rho)^2$. This form of the Virasoro Ward
identity is almost identical to the form of the Virasoro Ward identity
on arbitrary Riemann surfaces as derived in \cite{ward},
although there one works with an holomorphic projective
connection and  the last term of (\ref{Virasoro3}) is therefore
absent.  The precise form of $\cite{ward}$ is recovered if one
takes an appropriate regularization of the induced action for
gravity, see appendix B.
The form of the Virasoro Ward-identity (\ref{Virasoro})
on the plane can be recovered by replacing $T$ by $T+
{\cal R}/2$, a fact already observed in \cite{egoo}.

Given this construction of $W$ algebras on an arbitrary Riemann
surface $\Sigma$,
it is straightforward to write down $W$ transformations
on a Riemann surface, and to construct the chiral and covariant
actions for $W$ gravity; one simply follows the construction in
sections~1 and 2, and replaces the WZW actions by their
generalization (\ref{wzw2}). The choice of 'base point' $B^0$ in
(\ref{wzw2}) is not really important, though one should realize
that one cannot in general take it to be equal to $\nabla$,
because $\nabla$ is not in general flat. However, when one
constructs the full covariant action, one will see that this
expression is independent of the choice of base-point $B^0$.
As an example of this procedure, we will in
Appendix B compute the covariant action for gravity once more,
but now on an arbitrary Riemann surface.

\newsubsection{The Moduli Space for $W$ Gravity}

The moduli space for $W$ gravity is in principle given by the
quotient of the space of $W$ lds by the space of
$W$ transformations. In our case the space of $W$ fields is
given by the set of operators $M=\{\nabla_z+\Lambda+W\}$. Each
operator $D'\in M$ defines an anti-holomorphic structure on the
bundle $V$ (\ref{def:V}). Such an anti-holomorphic structure is
determined by defining what the local anti-holomorphic sections
of the vector bundle are. In the anti-holomorphic structure
corresponding to $D'$ these are just the local sections $s$ that
satisfy $D's=0$. The space of anti-holomorphic structures $M$
must be divided by the set of $W$ transformations. To do this,
introduce the following equivalence relation on $M$: two
operators $D'_1,D'_2 \in M$ are equivalent, $D'_1 \sim D'_2$, if
there is a gauge transformation $g\in {\cal G}_c$ relating the
two, $D'_1=(D'_2)^g$. The moduli space we are looking for
is the space ${\cal M}_W=M/\sim$. The transformations that relate
two different $D'$ are what one might call global
$W$ transformations. The infinitesimal transformations of this
type are precisely the $W$ transformations considered
previously. Note that the equivalence relation $D'_1\sim D'_2$
is not generated by the action of a group on $M$, as the precise
form of the gauge transformation relating two different $D'$
depends explicitly on the precise form of these $D'$. Thus, we
cannot view ${\cal M}_W$ as the quotient of some space by a group
action, and this makes the study of ${\cal M}_W$ somewhat more
difficult. One of the things we would in particular like to
compute is the dimension of the ${\cal M}_W$, or equivalently, of
its tangent space. If one were to consider the full set of
anti-holomorphic structures on $V$ modulo gauge transformations,
\ie \ the space ${\cal M}=\{ \nabla_z + A_z \}/{\cal G}_c$,
the tangent space  $T_{D'}{\cal M}$ at $D'\in {\cal M}$ is given
by the $(1,0)$-cohomology of the short complex
\be \label{complex1}
0\pijl \Omega^0 (\Sigma;\add(P_c))
\pijl \Omega^{1,0} (\Sigma;\add(P_c)) \pijl 0.
\ee
Here, $\Omega^{p,q}(\Sigma;\add(P_c))$ denotes the space of
$(p,q)$-forms with values in $\add(P_c)$.
To compute the tangent space $T_{D'}{\cal M}_W$ for ${\cal
M}_W$, we should replace this complex
by some kind of $W$ complex containing the
$W$ transformations. There is an interesting connection between
the two, which we will now explain. This connection relies
heavily on the existence of the operator $L$ that was defined as
the inverse of $\ad{\Lambda}$ in section~2. Because $L$ is an
operator of degree $-1$, it provides us with an `integration'
operator
\be \label{integr}
 \Omega^{1,0} (\Sigma;\add(P_c))
 \stackrel{L}{\longrightarrow}
 \Omega^0 (\Sigma;\add(P_c)).
 \ee
 As an analogy one might think of the operation of integrating
 over the $n^{th}$ co-ordinate in $\re^n$, which maps $p$-forms
 on $\re^n$ to $(p-1)$-forms on $\re^{n-1}$. This latter
 operator can be used to show that the cohomology of $\re^n$ is
 the same as the cohomology of $\re^{n-1}$, by constructing a
 so-called homotopy-equivalence between the de Rham complexes for
 $\re^n$ and $\re^{n-1}$ \cite{botttu}. Here we can perform a
 similar construction using the `integration' operator $L$.
 Defining the two operators
 \be \label{def;f}
 f_0=1-L\circ D' ,\,\,\,\,\,\, f_1=1-D' \circ L,
 \ee
 we can construct the following commutative diagram
 \be \label{complex2}
\begin{array}{rcccl}
0\pijl &
 \Omega^0 (\Sigma;\add(P_c)) & \pijl &
 \Omega^{1,0} (\Sigma;\add(P_c)) & \pijl 0 \\
 & \Bigl \downarrow f_0 & & \Bigl\downarrow f_1 & \\
0\pijl &
 f_0(\Omega^0 (\Sigma;\add(P_c))) & \pijl &
 f_1(\Omega^{1,0} (\Sigma;\add(P_c))) & \pijl 0
 \end{array}
 \ee
 which gives actually a homotopy equivalence of complexes\footnote{
 Indeed, denoting by $f$ both maps $f_0$ and $f_1$, we have
 $1-f=L\circ D' + D' \circ L$, so that $L$ is
 precisely an homotopy operator as defined in \cite{botttu}.},
 implying that the cohomology of both complexes in
 (\ref{complex2}) is the same. The next step is to iterate this
 construction a number of times, until the complex does not
 change anymore. Let us denote the corresponding limit complex,
 if it exists, by
 \be \label{complex3}
0\pijl f_0^{\infty}(\Omega^0 (\Sigma;\add(P_c)))
\pijl f_1^{\infty}(\Omega^{1,0} (\Sigma;\add(P_c)))
\pijl 0.
\ee
Using the properties of $L$ one can show that a sufficient
condition for the limit complex to exist is that the operator
$L\circ(D'-\ad{\Lambda})$ is nilpotent, and that in that
case
\ba
f_0^{\infty} = (1-L\circ D')^{\infty} & = &
\frac{1}{1+L(D'-\ad{\Lambda})} \circ \prk, \nonu
f_1^{\infty} = (1-D'\circ L)^{\infty} & = & \prkd \circ
\frac{1}{1+(D'-\ad{\Lambda})L} . \label{def:f}
\ea
Specializing to the case of $W$ algebras, we take
$D'=\nabla_z+\Lambda+W$ and find, upon comparing
the limit complex with (\ref{eq:abar4}), that the limit complex
precisely contains the $W$ transformations, and is the
$W$ complex we were looking for. To illustrate how this works in
practice, we take again $G=Sl(2,\re)$ as we did in section~5.3. The
limit complex is reached by applying $f_0$ and $f_1$ two times.
The operator $L$ is given by
\be \label{def:L}
L:\mats{p^0}{p^+}{p^-}{-p^0}=\mats{-p^+/2}{0}{p^0}{p^+/2},
\ee
and if we represent an arbitrary element of
$\Omega^0 (\Sigma;\add(P_c))$ by $\mats{\eps^0}{\eps^+}{
\eps^-}{-\eps^0}$, and an element of
$\Omega^{1,0} (\Sigma;\add(P_c))$ by
$\mats{a^0}{a^+}{a^-}{-a^0}$, we find the following diagram,
where $\gamma=\dif\log\rho$:
\bd \label{complex4}
\begin{array}{ccc}
\mats{\eps^0}{\eps^+}{\eps^-}{-\eps^0} & \pijl &
\mats{a^0}{a^+}{a^-}{-a^0}  \\
  \Bigl\downarrow f_0 & & \Bigl\downarrow f_1  \\
\mats{\hf\dif\eps^++\hf\eps^+\gamma}{\eps^+}{\eps^+
T-\dif\eps^0}{-\hf\dif\eps^+-\hf\eps^+\gamma} & \pijl &
\mats{\hf\dif a^+}{0}{a^--(\dif-\gamma)a^0+Ta^+}{-\hf\dif
a^+} \\
  \Bigl\downarrow f_0 & & \Bigl\downarrow f_1  \\
\mats{\hf\dif\eps^++\hf\eps^+\gamma}{\eps^+}{\eps^+
T-\hf\dif^2\eps^+-\hf\dif(\eps^+\gamma)
}{-\hf\dif\eps^+ -\hf\eps^+ \gamma} & \pijl &
\mats{0}{0}{a^--(\dif-\gamma)(a^0+(\hf\dif-T)a^+)}{0}
 \end{array}
\ed
Working out the action of $D'$ in the last line we find
\be \label{trafo2}
D'
\mats{\hf\dif\eps^++\hf\eps^+\gamma}{\eps^+}{\eps^+
T-\hf\dif^2\eps^+-\hf\dif(\eps\gamma)
}{-\hf\dif\eps^+-\hf\eps^+\gamma} =
\mats{0}{0}{\delta_{\eps^+} T}{0},
\ee
where $\delta_{\eps^+}T =-\hf(\dif-\gamma)\dif(\dif+\gamma)\eps^+ +
2\dif\eps^+ T + \eps^+\dif T$
which indeed describes the transformation of $T$ under
a co-ordinate transformation.

Altogether we reach the remarkable
conclusion that $W$ transformations are nothing but a homotopic
contraction of ordinary gauge transformations. Under a homotopy
equivalence the cohomology does not change, and therefore the
Riemann-Roch theorem can be applied to the $W$ complex
(\ref{complex3}) to give
\be \label{eq:rr}
\dim H^{1,0} - \dim H^{0,0} = (g-1) \dim G = (g-1)(N^2-1).
\ee
This is a useful formula which we need to prove the following:
for genus $g>1$, ${\cal M}_W=M/\sim=M^{hol}/\sim$, where
$M^{hol}=\{\nabla_z+\Lambda+W \mid \bnab W=0 \}$. In
other words, the fields $W$ can always be made holomorphic using
a global $W$ transformation. To prove this, it is sufficient
to show that if we write down an even further reduced complex
containing $D'\in M^{hol}$ and only those $W$ transformations
that preserve the condition $D'\in M^{hol}$, this complex still
has the same cohomology as (\ref{complex3}). It might happen
that in this way one misses certain connected components of
${\cal M}_W$, but that is not a problem here:
${\cal M}={\cal M}_W$ is connected, because we are working with
bundles of a fixed topological type.

The infinitesimal gauge transformations that preserve the condition
$\bnab W=0$, are given by the $\eps$ satisfying
\be \label{cond1}
\bnab (\nabla_z + \ad{\Lambda} + \ad{W}) \eps=0.
\ee
If we choose a metric of constant curvature
$R_{z\bar{z}}=[\nabla_z,\bnab]$, then
$L(R_{z\bar{z}})$ is proportional to the Lie-algebra element
$\Lambda^-$ that defines a highest weight gauge (\ref{eq:sl2}).
This shows that $[L(R_{z\bar{z}}),W]=0$ and
\be \label{holward}
[\nabla_z+\Lambda+W,\bnab-L(R_{z\bar{z}})]=0
\ee
for the $W$ that satisfy $\bnab W$=0. Note that
(\ref{holward}) gives a solution to the zero-curvature equations
for these $W$.  The $\eps$ that satisfy (\ref{cond1}) must also
be of the form $\eps=(1+L(\nabla_z+\ad{W}))^{-1} F$ with
$F\in\prk\bg$ in order to preserve the form of $W$.
If we substitute this in (\ref{cond1}) and use the fact that
$[L(R_{z\bar{z}}),\delta W]=0$, (\ref{cond1}) can be rewritten
as
\ba
(\bnab-L(R_{z\bar{z}}))(\nabla_z+\ad{\Lambda}+\ad{W})
\frac{1}{1+L(\nabla_z+\ad{W})} F & = & 0 \Leftrightarrow \nonu
(\nabla_z+\ad{\Lambda}+\ad{W}) (\bnab-L(R_{z\bar{z}}))
\frac{1}{1+L(\nabla_z+\ad{W})} F & = & 0 \Leftrightarrow \nonu
(\nabla_z+\ad{\Lambda}+\ad{W})
\frac{1}{1+L(\nabla_z+\ad{W})} \bnab F & = & 0 .
\label{cond2}
\ea
Locally, $\bnab F$ can be written as $\sum_{\alpha}
f_{\alpha}(\bar{z}) G_{\alpha}(z)$, where the $f_{\alpha}$ are
linearly independent antiholomorphic functions, and the
$G_{\alpha}$ are holomorphic sections with respect to
$\bnab$ of $(\prk\add(P_c))\otimes \bar{K}$. Substituting
this in (\ref{cond2}) yields
\be \sum_{\alpha} f_{\alpha}
(\nabla_z+\ad{\Lambda}+\ad{W})
\frac{1}{1+L(\nabla_z+\ad{W})} G_{\alpha}  =  0 .
\label{cond3}
\ee
Because the $f_{\alpha}$ are linearly independent, each $G_{\alpha}$
must satisfy
$(\nabla_z+\ad{\Lambda}+\ad{W})(1+L(\nabla_z+\ad{W}))^{-1}
G_{\alpha}=0$. Locally, there are a finite number of solutions
$G_{\alpha}$ to this equation.
Globally, such $G_{\alpha}$ do not exist, as
$(\prk\add(P_c))\otimes\bar{K}$ is a direct sum of line bundles
$K^r$ with $r<0$ (upon identifying $\bar{K}$ with
$K^{-1}$), and these do not have any global holomorphic
sections. Therefore (\ref{cond2}) implies that
$\bnab F=0$. $F$ is a section of a direct sum of line bundles $K^r$
with $r\leq 0$. These do, for genus $g>1$,
not have global holomorphic sections
unless $r=0$, in which case the only holomorphic sections are the
constant ones. The piece of $F$ which transforms as a section of
$K^0$ is precisely the piece that has degree zero with respect to
the gradation of the Lie algebra. Let us denote the subalgebra
in which this piece of $F$ lives by $\bg_0=\prk\prkd\bg$\footnote{That
$\bg_0$ is actually a subalgebra is related to the fact that
$\bg_0$ contains the Kac-Moody symmetries that survive the
reduction to the $W$ algebra. One can in principle impose
further constraints on the $W$ algebra so as to get rid of these
residual Kac-Moody symmetries \cite{wlnpol}, but we will not do
that here.}. For a constant $F\in\bg_0$ the parameter $\eps$ of
the gauge transformation is given by
$\eps=(1+L(\nabla_z+\ad{W}))^{-1} F=F$, and the gauge
transformation reads $\delta W=[W,F]$. The reduced complex we
were looking for is
\be \label{complex5}
0\pijl \bg_0 \pijl T_{D'}M^{hol} \pijl 0.
\ee
To show that the cohomology of this complex agrees with that of
(\ref{complex3}) we need only compute the difference $\dim
H^{1,0}-\dim H^{0,0}$ of (\ref{complex5}). In (\ref{complex5})
only finite dimensional spaces occur, and therefore the index
$\dim H^{1,0} - \dim H^{0,0}$ equals $\dim T_{D'}M^{hol}-\dim
\bg_0$. The dimension of $M^{hol}$ equals $\sum_i
H^0_{\bnab}(\Sigma;K^{s_i})$, where $s_i$ are the
spins of the different components of $W$. The dimension of
$H^0_{\bnab}(\Sigma;K^r)$ equals $(2r-1)(g-1)$ for
$r>1$, and $g$ for $r=1$. Thus we find
\ba
\dim H^{1,0}-\dim H^{0,0} & = & \sum_{i,s_i>1}(g-1)(2 s_i-1) +
\sum_{i,s_i=1} g - \dim \bg_0 \nonu
& = & \sum_i (g-1)(2s_i-1) = (g-1)\dim G, \label{index2}
\ea
which indeed agrees with (\ref{eq:rr}). In the last line we used
(\ref{spinid}). Altogether this proves that ${\cal
M}_W=M^{hol}/\sim$, and so we have a simple finite dimensional
model of $W$ moduli space at our disposal.

Although the dimensions computed so far strongly hint that $W$ moduli
space has something to do with the moduli space of
$Sl(N,\re)$-bundles, it is at this stage not clear what the
precise relation, if it exists, should be. The zero-curvature
equations associate a flat connection to all operators
$D'=\nabla_z+\ad{\Lambda}+\ad{W}$, but a priori these flat
connections are flat $Sl(N,\ce)$ connections, and it is not easy
to see whether they can be written as flat $Sl(N,\re)$ connections
using an appropriate gauge transformation. The main difficulty is
that $Sl(N,\re)$ is a non-compact group, and therefore one
cannot simply use the Narasimhan-Seshadri theorem
\cite{moduli2} (see also \cite{moduli}),
which essentially states that for compact groups
the space of anti-holomorphic structures on an associated vector
bundle modulo complexified gauge transformations is the same as
the space of flat connections modulo ordinary gauge
transformations. In this theorem, the anti-holomorphic structure
is required to satisfy a certain condition called stability, and
this condition is not valid for the special bundles under
consideration.

There exists an extension of the work of Narasimhan-Seshadri
where the compact group is replaced by the general linear group
$Gl(N,\ce)$. This is the theory of Higgs bundles
\cite{hitchin,simpson}, and this seems to be the natural
setting for $W$ moduli space. A Higgs bundle is a pair
consisting of a holomorphic vector bundle $V$ and a holomorphic
section $\theta\in H^0(\Sigma;\End(V)\otimes K)$. In our case we
are interested in the situation where $V$ is given by
(\ref{def:V}), the holomorphic structure is given by the
operator $\bnab$, the group $Gl(N,\ce)$ is reduced to
$Sl(N,\ce)$ and $\theta=\Lambda+W$, where $W$
is holomorphic. The group
$Gl(N,\ce)$ acts in a natural way on Higgs bundles, and one can
define a moduli space for Higgs bundles by identifying two that
are equivalent under a $Gl(N,\ce)$-transformation. To obtain a
good moduli space one has to impose a condition on the Higgs
bundle that is also called stability. A Higgs-bundle is called
stable if for every holomorphic subbundle $V'\subset V$ that
satisfies $\theta(V')\subset V'\otimes K$, the slope $\mu(V')$
of $V'$ is smaller than the slope $\mu(V)$ of $V$. The slope is
defined as the first Chern class divided by the rank of the
bundle.

Let us see whether the Higgs bundle with
$\theta=\Lambda+W$ is stable. The slope of $V$ vanishes, and
therefore every subbundle $V'$ with $\theta(V')\subset V'\otimes
K$ must have a negative slope for stability.
The $sl_2$-algebra (\ref{eq:sl2}) acts via left multiplication on
the vector bundle $V$. Under this action the $N$-dimensional
representation furnished by $V$ decomposes in a direct sum of
irreducible $sl_2$-representations, $V=\bigoplus_{l=1}^{n_l}
V_l$, of spin
$j_l$, and $V_l\simeq K^{-j_l}\bigoplus K^{1-j_l}\bigoplus
\ldots\bigoplus K^{j_l}$. The slope of each of the
$V_l$ is zero, as they have vanishing Chern class.
$\Lambda$ preserves $V_l$ and all subbundles of
$V_l$ of the type $K^{-j_l}\bigoplus K^{1-j_l} \bigoplus \ldots
\bigoplus K^{j_l-t}$ for some $t>0$.
These all have strictly negative slope,
and therefore the only problematic subbundles of $V$ are direct
sums of the $V_l$, as these are the only holomorphic
subbundles preserved by $\Lambda$
that have a nonnegative slope. The same bundles are
also the bundles that might threaten the stability of $(V,\theta)$
with $\theta=\Lambda+W$.
Now there is a component of $W$ for every irreducible
$sl_2$ representation in $(\bigoplus V_l)\otimes(\bigoplus
V_l)$, except for one overall trivial representation. A
component $W$ corresponding to an irreducible subrepresentation of
$V_l\otimes V_{l'}$ mixes between the bundles $V_l$ and
$V_{l'}$. Therefore if sufficiently many of these components are
nonzero, no direct sum of $V_l$'s will be invariant under
$\Lambda+W$ anymore, and all proper holomorphic
subbundles, if they exist, will have negative slope.

Another, equivalent way to express this condition is to demand that
$\ker\ad{W} (\bg_0)=0$, so that $\bg_0$ acts faithfully on
$M^{hol}$. If we therefore define
$M^{hol}_{red}=\{\nabla_z+\Lambda+W\mid \bnab W =0 \wedge
\ker\ad{W}(\bg_0)=0\}$, then the quotient space
$M^{hol}_{red}/\bg_0$ has no singularities, and it is
naturally a subspace of the moduli space of stable Higgs
bundles, of complex dimension $(g-1)\dim(G)$.

The dimension of the moduli space of Higgs bundles is
$2(g-1)\dim(G)$, which is twice as large as the dimension of the
$W$ moduli space. These correspond to flat irreducible
$Sl(N,\ce)$ bundles
over the Riemann surface $\Sigma$ \cite{hitchin,simpson2}.
One might wonder which property characterizes the flat $Sl(N,\ce)$
connections that correspond to points in the $W$ moduli space.
For general $W$ algebras we do not know the answer to this
question, but for the `standard' $W$ algebras
the answer is, that only those flat $Sl(N,\ce)$ connections which
are reducible to a flat $Sl(N,\re)$ connections can correspond to
points in the $W$ moduli space.
To prove this, we use
lemma 3.20 in \cite{simpson}. This lemma states that a Higgs
bundle $(V,\theta)$ corresponds to a flat real connection if and
only if there exists a bilinear symmetric form $S(u,v)$
on $V \otimes V_C$, where $V_C$ is the Higgs bundle
$(V,-\theta)$, such that
\be \label{def:r}
\dbar S(u,v)=S((\bnab+\theta)u,v)+
S(u,(\bnab-\theta)v).
\ee
For the standard $W$ algebras
such a symmetric form $S$ exists. The vector bundle $V$ is in
this case
\be
\label{decomp}
V=\bigoplus_{l=1}^{N} V_l, \,\,\,\,\,\,\,\, V_l\simeq
K^{\frac{l-N-1}{2}}
\ee
and the symmetric form $S$ is given by
\be
\label{def:S}
S(u,v)=\sum_{l=1}^N u_l \,\,\, v_{N+1-l}.
\ee
It is an easy exercise to
show that (\ref{def:r}) indeed holds for (\ref{def:S}). Putting
everything together we conclude that, for standard
$W$ algebras, $W$ moduli space is a
component of the moduli space of flat irreducible $Sl(N,\re)$
connections. This parametrization of a component
of the moduli space of flat
$Sl(N,\re)$ connections in terms of certain Higgs bundles
is similar to the one mentioned by Hitchin \cite{hitchin2}.
The relevant component is specified by the
topological type of the real vector bundle on which the flat
$Sl(N,\re)$ connection lives. To really construct this flat
connection explicitly, one needs to know the so-called
Hermitian-Yang-Mills metric on the Higgs bundle, see
\cite{simpson2}. This metric, and the associated flat connection
are very easy to describe if $W=0$. In that case one picks a
constant curvature metric on the Riemann surface, and uses the
metric this induces on $K$ to construct a metric on $V$. This
is already the Hermitian-Yang-Mills metric and the corresponding
flat connection is
\be \label{eq:flat}
D=\nabla_z+\bnab+\Lambda-L(R_{z\bar{z}}).
\ee
The real vector bundle for which this defines an $Sl(N,\re)$
connection is given by the bundle left invariant by an
involution of $V$ that commutes with $D$. The involution
$\sigma:V\rightarrow V$ is given by sending $u\in K^r
\rightarrow \bar{u}\in \bar{K}^r \simeq K^{-r}$, where the
metric is used to identify $\bar{K}$ with $K^{-1}$.

As an example, consider ordinary gravity. In that case $G=Sl(2,\re)$,
and $V=K^{-\hf} \bigoplus K^{\hf}$. The involution $\sigma$ is
in local co-ordinates with metric $ds^2=\rho dz d\bar{z}$
given by
\be \label{def:inv}
\sigma :\left(
\begin{array}{c} u_1 \\ u_2 \end{array}
\right) \rightarrow \left(
\begin{array}{c} \bar{u}_2/\sqrt{\rho} \\
\bar{u}_1 \sqrt{\rho} \end{array} \right).
\ee
The corresponding real bundle is one of Euler class $2(g-1)$,
and the $W$ moduli space as defined here is the
Teichm\"uller space of $\Sigma$, which is indeed closely related
to the moduli space of Riemann surfaces, and has occurred before
in studies of 2D quantum gravity \cite{herman}. For details, see
\cite{hitchin}.

It is a very interesting problem to characterize the flat
$Sl(N,\ce)$-bundles that are related to the moduli space of
the nonstandard $W$ algebras. We have checked for a few cases
that it is impossible to construct a symmetric
bilinear form satisfying (\ref{def:r}) for those $W$ algebras,
and therefore they do not correspond to flat $Sl(N,\re)$
connections. If we replace $V$ by $Sl(V)$, and consider the
Higgs bundle $(Sl(V),\theta)$ with $\theta:Sl(V)\rightarrow
Sl(V)\otimes K$ given by $\theta(X)=[\Lambda+W,X]$, then the
existence of symmetric bilinear form satisfying
(\ref{def:r}) for this Higgs bundle would show that the
flat $Sl(N,\ce)$ connections for these $W$ algebras are always
reducible to a flat $\bg$ connection, where $\bg$ is some Lie
algebra whose complexification is $sl(N,\ce)$. However,
for the cases we checked it was impossible to construct a
symmetric form $S$ for $(Sl(V),\theta)$ either, and therefore
it is still unclear what precisely characterizes the
nonstandard $W$ moduli spaces.

Instead of looking at $M^{hol}_{red}/\bg_0$, one could also look at
different `strata' of $M^{hol}$, by defining $M^{hol}_k=
\{\nabla_z+\Lambda+W\mid \bnab W =0 \wedge
\dim \ker\ad{W}(\bg_0)=k\}$. The space ${\cal
M}_{W,k}=M^{hol}_k/\bg_0$ is presumably related to `singular'
configurations of $W$ fields, and deserves some further study
as well.

The simplest nonstandard $W$ algebra is $W_3^{(2)}$ \cite{bersch}.
For this $W$ algebra, $\Lambda$ and $W$ are given by
\be \label{def:lw}
\Lambda=\mat{0}{0}{1}{0}{0}{0}{0}{0}{0}, \,\,\,\,\, W=
\mat{J}{0}{0}{G^+}{-2J}{0}{T}{G^-}{J},
\ee
where $J$ has spin 1, $G^+,G^-$ have spin $3/2$ and $T$ has spin
$2$. The space $M^{hol}$ has dimension $8g-7$, and $\bg_0$ consists
of the constant matrices
\be \label{trafow32}
X=\mat{\eps}{0}{0}{0}{-2\eps}{0}{0}{0}{\eps}, \,\,\,\, \mbox{\rm
and}\,\,\,\,\,\,
\delta_{\eps}W=[W,X]=\mat{0}{0}{0}{3\eps G^+}{0}{0}{0}{-3
\eps G^-}{0}.
\ee
{}From this we see that $M^{hol}_{red}=\{J,G^+,G^-,T | G^+\neq 0
\,\,\,\,\mbox{\rm or}\,\,\,\, G^-\neq 0\}$, and that ${\cal M}_{W,0}$ is
topologically the product of $\ce^{4g-3}$ and a weighted projective
space of dimension $4g-5$. Clearly, ${\cal M}_{W,1}$ is
topologically a vector space of dimension $4g-3$. We see that for
this $W$ algebra the moduli space is non trivial.

The discussion of $W$ moduli space in this section has so far
been limited to genus $g>1$. Most of the analysis can also be
carried through for genus $g=0,1$. The main difference with
$g>1$ is that for the latter case, $W$ moduli space is in a
natural way a subspace of the moduli space of stable Higgs
bundles. For $g=0,1$ this is no longer the case, because the
Higgs bundles one obtains for $g=0,1$ are not stable any more,
the reason for this being the fact that the first Chern class
of $K$ is given by $c_1(K)=2(g-1)$, and changes sign at $g=1$.
Let us briefly indicate what the moduli spaces for $g=0,1$ look
like.

For $g=0$, the line bundle $K^r$ with $r>0$ has no global
holomorphic sections. Therefore $M^{hol}$ is contains only
$D'=\nabla+\Lambda$ and has dimension 0. The gauge transformations
that act trivially on $\nabla+\Lambda$ are given by
$\delta\Lambda=0=[\Lambda,F]$, where $F$ is an arbitrary
holomorphic section of $\prk\add(P_c)\simeq\oplus_i K^{1-s_i}$.
For genus $g=0$ the line bundle $K^{1-s_i}$ has $(2s_i-1)$
holomorphic sections, and the dimension of the space of gauge
transformations that act on $M^{hol}$ equals
$\sum_i(2s_i-1)=\dim G$. This shows that $\dim H^{1,0}-\dim
H^{0,0}=-\dim G$, in agreement with the Riemann-Roch theorem
(\ref{complex3}), which is valid for arbitrary genus.

For $g=1$, the line bundles $K^r$ are all trivial and have
precisely one holomorphic section. If $d_W$ denotes the number
of generators of the $W$ algebras, then $\dim M^{hol}=d_W$, and
the space of gauge transformations that acts on $M^{hol}$ also
has dimension $d_W$. These gauge transformations act on
$M^{hol}$ via $\delta W=[\Lambda+W,(1+L \add_W)^{-1} F]$, where
$F$ is an arbitrary holomorphic section of $\prk\add(P_c)$.
Again, (\ref{complex3}) is satisfied. A natural candidate for
the moduli space is in this case
\be \label{torus}
{\cal M}_W=\{\nabla+\Lambda+W | \dim
C_{\bg}(\Lambda+W)=\mbox{\rm rank } \bg \}/\sim~,
\ee
where $C_{\bg}(X)$ is the centralizer of $X$, \ie\ the set of
elements of $\bg$ that commute with $X$, and $\bg$ should be
identified with the holomorphic sections of $\add(P_c)\otimes
K$. The dimension of the genus 1 moduli space equals the rank of
$\bg$.

Actually, what we really have been computing up till now is $W$
Teichm\"uller space rather than $W$ moduli space. For ordinary
gravity, the latter is given by the quotient of the former by the
action of the modular group.
This suggests that a good candidate for $W$
moduli space is also to consider
the quotient of the $W$ Teichm\"uller space by
the action of the modular group (cf. \cite{li2}). It is an interesting
problem to investigate these spaces in some more detail, and to
try to generalize the framework for ordinary gravity
coupled to matter, where one expresses correlation functions
in terms of modular invariant combinations of conformal blocks,
to the case of matter theories coupled to $W$ gravity.

\newpage

\noindent {\bf Acknowledgement}\\[7mm]
\noindent We would like to thank Jaap Kalkman for stimulating
discussions and useful comments. This work was financially
supported by the Stichting voor Fundamenteel Onderzoek der Materie
(FOM).

\startappendix

\newsection{Appendix}

In this Appendix we give the definitions and some of the properties
of the WZW actions $S^{\pm}_{wzw}$:
\be \label{deff}
S^{\pm}_{wzw}(g)=\deel{1}{4\pi}\int_{\Sigma}d^2z\,\tr(g^{-1}\dif
gg^{-1}\dbar g)\pm \deel{1}{12\pi}\int_B \tr(g^{-1}dg)^3,
\ee
satisfying the following Polyakov--Wiegman identities \cite{powie}:
\ba \label{idd}
S_{wzw}^{+}(gh) &=& S_{wzw}^{+}(g)+S_{wzw}^{+}(h)
+\act{g^{-1}\dbar g\dif h h^{-1}}, \nonu
S_{wzw}^{-}(gh) &=& S_{wzw}^{-}(g)+S_{wzw}^{-}(h)
+\act{g^{-1}\dif g \dbar h h^{-1}}.
\ea
The equations of motion resulting from these WZW actions were
already studied in section~5.2 (see (\ref{varwzw2})). If we view
$S^{-}_{wzw}(g)$ as a function of $A_z=g^{-1}\dif g$, the
equation of motion is:
\be \label{bvgl}
2\pi \vars{S^{-}_{wzw}(A_z)}{A_z}=\jbar=0,
\ee
where $\jbar$ is such that the pair $(A_z,\jbar)$ have vanishing
curvature, so $\jbar=g^{-1}\dbar g$.
Equivalently, if we take $S^{+}_{wzw}(h)$ to be a
function of $\abar=h^{-1}\dbar h$, we have:
\be \label{bvgl2}
2\pi \vars{S^{+}_{wzw}(\abar)}{\abar}=J_z=0,
\ee
where now $F(J_z,\abar)=0$. For the action
$kS^{+}_{wzw}(h)-\act{\Lambda h^{-1}\dbar h}$, where $(h^{-1}\dbar
h)^+$ is held fixed and
should be identified with $\mu$, this implies that
the equations of motion can only be satisfied if
$h^{-1}\dif h$ is of the form:
\be \label{formagain}
h^{-1}\dif h= \left(
\begin{array}{cc} 0 & 1 \\ {*} & 0
\end{array} \right).
\ee
{}From section~3 we know that $*$ should be identified with a
spin two field, and that the solution for $\abar=h^{-1}\dbar h$
is now that $\abar$ is as in (\ref{defabar}), since this is the
unique answer such that (\ref{formagain}) and
$\abar$ have vanishing curvature, as we argued in section 2.2.

\newsection{Appendix}

In this appendix we will sketch how some of the results of
section~2 generalize to arbitrary Riemann surfaces, using the
generalized WZW functional (\ref{wzw2}). The chiral action
$\Gamma[T]$ was in section~2 shown to be given by $-k
S^-_{wzw}(g)$, where $g^{-1}\dif g=\Lambda+W$. This has a
natural generalization, namely $\Gamma[T]=-kS^-_{wzw}(A;B)$,
where $A$ is the flat connection consisting of (\ref{eq:hol})
and (\ref{eq:ahol}), and $B$ is a {\it fixed} flat reference
connection, representing a particular regularization, as
explained in section~5.2. If $B$ locally has the form
\be \label{def:b}
B=\mats{B_z^0}{B_z^+}{B_z^-}{-B_z^0} dz +
\mats{B_{\bar{z}}^0}{B_{\bar{z}}^+}{B_{\bar{z}}^-}{-B_{\bar{z}}^0}
d\bar{z} ,
\ee
then the Ward-identity satisfied by $\Gamma[T]$ is given by
(\ref{Virasoro3}), with $\mu$ replaced by $B_{\bar{z}}^+
-\deel{2\pi}{k}\vars{\Gamma[T]}{T}$. By Fourier transformation,
we define the induced action $\gam[\mu]$
\be \label{ft2}
\exp(-\gam[\mu]) =
\int DT \exp\left( -\gam[T]-\deel{k}{2\pi}\int
d^2z\,(\mu-B_{\bar{z}}^+)(T-B_z^-) \right) ,
\ee
which differs from (\ref{ft}) by the explicit appearance of $B$
in the definition of Fourier transformation. If we define
$T(\mu)$  by requiring the connection
(\ref{eq:hol})+(\ref{eq:ahol}) to be flat, \ie\ we view $A$ as a
flat connection depending on $\mu$ rather than $T$, then in a
saddle point approximation the action $\gam[\mu]$ is simply
given by
\be \label{def:gmu}
\gam[\mu]= \gam[T(\mu)]+\deel{k}{2\pi}\int
d^2z\,(\mu-B_{\bar{z}}^+)(T(\mu)-B_z^-) .
\ee
The 'induced' action $\gam[\mu]$ satisfies the Ward identity
\be \label{Virasoro5}
(\dbar-\mu\dif-2(\dif\mu))
\vars{\gam[\mu]}{\mu}=\deel{c}{24\pi}(\dif^3+2{\cal R'}\dif+(\dif{
\cal R'}))\mu-\deel{c}{24\pi}\dbar{\cal R'},
\ee
where ${\cal R'}=\dif^2
\log\rho-\hf(\dif\log\rho)^2-2B_z^-$. If we choose $B_z^-$ in
such a way so that ${\cal R'}$ is a holomorphic projective
connection, then we find precisely the Ward identity on a
general Riemann surface as found in \cite{ward}. The induced
action $\gam[\mu]$ on arbitrary Riemann surfaces has been
studied from different points of view in \cite{chirac}.

To construct the covariant action, we also need the left-moving
sector of the theory, \ie\
the actions $\gam[\bar{T}]$ and
$\gam[\mub]$. To construct these, we need to impose constraints
on the $\bar{z}$-component of a connection $A_2$. Using the
isomorphism between $\bar{K}$ and $K^{-1}$, one finds that one
has to impose the following constraint on $A_{2,\bar{z}}$
\be \label{constr2}
\bar{D}_{A_{2,\bar{z}}}=\dbar+\mats{0}{\bar{T}/\rho}{\rho}{0},
\ee
and
\be \label{constr3}
D_{A_{2,z}}=\dif+\mats{*}{*}{\mub\rho}{*}
\ee
is such that $A_2$ has vanishing curvature. Now
$\gam[\bar{T}]=-kS^-_{wzw}(B;A_2)$, and $\gam[\mub]$ is the
Fourier transform of $\gam[\bar{T}]$. In order to able to
construct a covariant action, we must use the same background
connection $B$ in both the left and right moving sector.
Repeating arguments similar to those in section~2, one finds
that the complete covariant action for gravity is simply given
by
\be \label{scov2}
S_{cov}(T,\bar{T},G)=-kS^-_{wzw}(A_1^G;A_2),
\ee
which does not depend on the choice of $B$ anymore, and is
manifestly invariant under both left and right diffeomorphisms.
To extract $\Delta\Gamma[T,\bar{T},G]$ from this covariant
action, we make repeatedly use of the generalized
Polyakov-Wiegmann identity (\ref{pw2}) to obtain
\ba
S_{cov}(T,\bar{T},G) & = & -kS^-_{wzw}(A_1;B)-kS^-_{wzw}(B^G;B)
-kS^-_{wzw}(B;A_2) \nonu
& & -\act{(A_1-B)_z (B-A_2^{G^{-1}})_{\bar{z}}} \nonu
& &  -\act{(A_1^G-B)_z (B-A_2)_{\bar{z}}} \nonu
 & & -\act{(A_1-B)_z G (A_2-B)_{\bar{z}} G^{-1} }. \label{scov3}
 \ea
In $-kS^-_{wzw}(A_1;B)$ and $-kS^-_{wzw}(B;A_2)$ we recognize
$\gam[T]$ and $\gam[\bar{T}]$, and the remainder of $S_{cov}$ in
(\ref{scov3}) is $\Delta\gam[T,\bar{T},G]$.
The `local counterterm' $\Delta\gam[\mu,\mub,G]$ can be obtained
from $\Delta\gam[T,\bar{T},G]$ by Fourier transformation, using
the same Fourier transformation as in (\ref{ft2}). If we
parametrize $G$ locally by the Gauss decomposition
\be \label{gauss3}
G=\mats{1}{0}{\om}{1}\mats{e^{\phi}}{0}{0}{e^{-\phi}}
\mats{1}{-\bom/\rho}{0}{1},
\ee
we find the following expression for $\Delta\gam[\mu,\mub,G]$ on
an arbitrary Riemann surface
\ba \label{cosl4}
& \Delta\gam[\mu,\mub,G]=\deel{k}{2\pi}\int d^2z\, \Bigl[
\dif \phi \dbar \phi + \omega(2\dbar \phi +(\dif+\dif\log\rho) \mu)
+\bar{\omega} ( 2\dif \phi + (\dbar+\dbar\log\rho) \bar{\mu}) & \nonu
&\hspace{2cm} +\mu \omega^2 + \bar{\mu}\bar{\omega}^2 +
2\omega\bar{\omega}-(1-\mu\bar{\mu})e^{-2\phi} -\phi\dif\dbar
\log\rho\Bigr] & \nonu
& + \,\,\mbox{extra piece}, &
\ea
where the extra piece contains the terms which depend on the
reference connection $B$:
\ba \label{cosl5}
\mbox{extra piece}&=&\act{B_z \mats{0}{-\mu}{\rho}{0}+
\mats{\dif\log\rho /2}{1}{-\mub\rho}{-\dif\log\rho/2}B_{\bar{z}} } \nonu
&-&  \act{B_z B_{\bar{z}}} + \deel{k}{\pi}\int d^2 z \, B_z^-
B_{\bar{z}}^+.
\ea
This expression can be further reduced by integrating out
$\omega$ and $\bar{\omega}$, and the result one obtains agrees
precisely with the local counterterms that have been constructed
previously in \cite{stora}, provided one chooses $B_z^-$ and
$B_{\bar{z}}^+$ in such a way that in the Ward-identity
(\ref{Virasoro5}) for $\gam[\mu]$ and in its analogue for
$\gam[\mub]$ only holomorphic and anti-holomorphic projective
connections occur. This shows that generalized WZW actions
provide a general and powerful framework to construct covariant
actions on arbitrary Riemann surfaces.

\end{document}